\documentclass[a4paper,11pt]{article}
\pdfoutput=1

\usepackage{amssymb,mathrsfs,amsmath}
\usepackage{a4wide}
\usepackage{color,xcolor}
\usepackage{slashed,soul}
\usepackage{graphicx}
\usepackage{amsfonts}
\usepackage{lscape}
\def\linkcolor{cyan!70!black}
\def\tabcolor{cyan!50!blue}
\usepackage[
colorlinks=true
,urlcolor=\linkcolor
,anchorcolor=\linkcolor
,citecolor=\linkcolor
,filecolor=\linkcolor
,linkcolor=\linkcolor
,menucolor=\linkcolor
,linktocpage=true
,pdfproducer=medialab
,pdfa=true
]{hyperref}
\usepackage{amsthm}
\usepackage{booktabs}
\usepackage{array}
\usepackage{rotating}
\usepackage[numbers, sort&compress]{natbib}
\usepackage{multirow}
\usepackage{float}
\usepackage[utf8]{inputenc}
\usepackage[T1]{fontenc}
\usepackage{appendix}

\newcommand{\eps}{\epsilon}
\newcommand{\ord}[1]{\mathcal{O}\left( #1 \right)}

\newcommand{\be}{\begin{equation}}
\newcommand{\ee}{\end{equation}}

\newcommand{\beq}{\begin{equation}} 
\newcommand{\eeq}{\end{equation}} 
\newcommand{\ba}{\begin{array}}  
\newcommand{\ea}{\end{array}} 
\newcommand{\bea}{\begin{eqnarray}}  
\newcommand{\eea}{\end{eqnarray} }  
\newcommand{\bal}{\begin{align}}
\newcommand{\eal}{\end{align}}   
\newcommand{\bi}{\begin{itemize}}  
\newcommand{\ei}{\end{itemize}}  
\newcommand{\ben}{\begin{enumerate}}  
\newcommand{\een}{\end{enumerate}}  
\newcommand{\bc}{\begin{center}}
\newcommand{\ec}{\end{center}} 
\newcommand{\bt}{\begin{table}}
\newcommand{\et}{\end{table}}  
\newcommand{\btb}{\begin{tabular}}
\newcommand{\etb}{\end{tabular}}

\newcommand{\fref}[1]{Figure~\ref{#1}} 
\newcommand{\eref}[1]{Eq.~(\ref{#1})}

\newcommand{\aref}[1]{Appendix~\ref{#1}}
\newcommand{\sref}[1]{Section~\ref{#1}}
\newcommand{\tref}[1]{Table~\ref{#1}}

\newcommand{\GeV}{\,\mathrm{GeV}}

\renewcommand{\Re}{{\rm Re}} 

\newcommand{\vp}{\varphi}
\newcommand{\sw}{s_{\mathrm{w}}}
\newcommand{\cw}{c_{\mathrm{w}}}

\def\Cone{\boldsymbol{C_1}}
\def\Ctwo{\boldsymbol{C_2}}
\def\Cthree{\boldsymbol{C_3}}
\def\Cfour{\boldsymbol{C_4}}
\def\mesonV{\mathcal V}
\def\mesonP{\mathcal P}

\newcommand\gVLV[1][\mesonV]{g_{VL}^{\tau\ell#1}} 
\newcommand\gVRV[1][\mesonV]{g_{VR}^{\tau\ell#1}} 
\newcommand\gTLV[1][\mesonV]{g_{TL}^{\tau\ell#1}} 
\newcommand\gTRV[1][\mesonV]{g_{TR}^{\tau\ell#1}} 

\newcommand\gLP[1][\mesonP]{g_{L}^{\tau\ell#1}} 
\newcommand\gRP[1][\mesonP]{ g_{R}^{\tau\ell#1}} 
\newcommand\gVLP[1][\mesonP]{ g_{VL}^{\tau\ell#1}} 
\newcommand\gVRP[1][\mesonP]{ g_{VR}^{\tau\ell#1}} 
\newcommand\gSLP[1][\mesonP]{ g_{SL}^{\tau\ell#1}} 
\newcommand\gSRP[1][\mesonP]{ g_{SR}^{\tau\ell#1}} 

\newcommand{\vpj }{\mbox{${\Phi^\dag\, \raisebox{1.5mm}{${}^\leftrightarrow$}\hspace{-4mm} D_\mu\,\Phi}$}}
\newcommand{\vpjt}{\mbox{${\Phi^\dag \tau^I\,\raisebox{1.5mm}{${}^\leftrightarrow$}\hspace{-4mm} D_\mu\,\Phi}$}}
\def\ocal{{\cal O}}
\def\lcal{{\cal L}}
\def\nn{\nonumber\\ }

\renewcommand{\baselinestretch}{1.2}

\newcommand{\lc}[1]{\textcolor{orange}{[LC: {#1}]}}

\allowdisplaybreaks

\begin{document}

\vspace{1cm}

\begin{titlepage}

\vspace*{-1.0truecm}
\begin{flushright}
TUM-HEP 1352/21 
 \end{flushright}
\vspace{0.8truecm}

\begin{center}
\renewcommand{\baselinestretch}{1.8}\normalsize
\boldmath
{\LARGE\textbf{
$Z$ lepton flavour violation as a probe for \\new physics at future $e^+e^-$ colliders
}}
\unboldmath
\end{center}

\vspace{0.4truecm}

\renewcommand*{\thefootnote}{\fnsymbol{footnote}}

\begin{center}

{\bf Lorenzo  Calibbi$\,^a$\footnote{\href{mailto:calibbi@nankai.edu.cn}{calibbi@nankai.edu.cn}}, 
Xabier Marcano$\,^b$\footnote{\href{mailto:xabier.marcano@tum.de}{xabier.marcano@tum.de}}
and Joydeep Roy$\,^{c}$\footnote{\href{mailto:joyroy.phy@gmail.com}{joyroy.phy@gmail.com}}}
\vspace{0.5truecm}

{\footnotesize

$^a${\sl School of Physics, Nankai University, Tianjin 300071, China \vspace{0.15truecm}}

$^b${\sl  Physik-Department, Technische Universit\"at M\"unchen, James-Franck-Stra\ss e, 85748 Garching, Germany \vspace{0.15truecm}}

$^c${\sl Center for High Energy Physics, Indian Institute of Science, Bangalore 560012, India \vspace{0.15truecm}}

}

\vspace*{5mm}
\end{center}

\renewcommand*{\thefootnote}{\arabic{footnote}}
\setcounter{footnote}{0}

\vspace{0.4cm}
\begin{abstract}
\noindent 
In this work we assess the potential of discovering new physics by searching for lepton-flavour-violating (LFV) decays of the $Z$ boson, $Z\to \ell_i \ell_j$, at the proposed circular $e^+e^-$ colliders CEPC and FCC-ee. Both projects plan to run at the $Z$-pole as a ``Tera Z factory'', {\it i.e.}, collecting $\ord{10^{12}}$ $Z$ decays.
In order to discuss the discovery potential in a model-independent way, we revisit  the LFV $Z$ decays in the context of the Standard Model effective field theory and study the indirect constraints from LFV $\mu$ and $\tau$ decays on the operators that can induce $Z\to \ell_i \ell_j$. 
We find that, while the $Z\to \mu e$ rates are beyond the expected sensitivities, a Tera Z factory is promising for $Z\to \tau\ell$ decays, probing New Physics at the same level of future low-energy LFV observables. 
\end{abstract}

\end{titlepage}

\tableofcontents

\section{Introduction}

The overwhelming evidence for neutrino oscillations has shown that lepton family numbers are not conserved and the Standard Model (SM) needs to be extended to account for neutrino masses. Hence, there is no fundamental reason why lepton flavour violation should not occur in processes involving charged leptons only, such as $\ell_i\to\ell_j \gamma$ ($i>j$). However, the rates of charged lepton-flavour-violating (LFV) processes induced
 by loops involving neutrinos and $W$ bosons are
  suppressed way below the sensitivity of any conceivable experiment by factors proportional to
 $\left( \Delta m^2/ M_W^2 \right)^2 \approx 10^{-49}$, where $\Delta m^2$ denotes the squared mass differences of 
  the neutrino mass eigenstates. As a consequence, LFV transitions are among the cleanest and most striking signals for physics beyond the SM (BSM), in fact beyond its minimal extensions accounting for neutrino masses. Fortunately, an intense experimental activity is ongoing, with a particular focus on low-energy LFV decays of leptons and mesons. For a review on LFV, including future experimental prospects, see~\cite{Calibbi:2017uvl}. 
  In this article, we will instead focus on a class of processes that is only accessible at high-energy colliders: lepton flavour violating decays of the $Z$ boson (LFVZD), $Z\to \ell_i \ell_j$, $i\neq j$. 
  
\begin{table}[t!]
\renewcommand{\arraystretch}{1.3}
\begin{center}
\begin{tabular}{cccc}
\hline
{Mode} & {LEP bound (95\% CL)} & {LHC bound (95\% CL)}  & {CEPC/FCC-ee exp.} \\
\hline
$\text{BR}(Z\to \mu e)$  &  $ 1.7\times 10^{-6}$\quad\cite{Akers:1995gz} & \hspace{-.55cm}  $ 7.5\times 10^{-7}$\quad\cite{Aad:2014bca} & $10^{-8}$\,--\,$10^{-10}$  \\
$\text{BR}(Z\to \tau e)$ &   $ 9.8\times 10^{-6}$\quad\cite{Akers:1995gz} &   $ 5.0\times 10^{-6}$\quad\cite{ATLAS:2020zlz,newATLAS} & $10^{-9}$ \\
$\text{BR}(Z\to \tau \mu)$ &  $ 1.2\times 10^{-5}$\quad\cite{Abreu:1996mj} &   $ 6.5\times 10^{-6}$\quad\cite{ATLAS:2020zlz,newATLAS}  & $10^{-9}$  \\
\hline
\end{tabular}
\caption{Current upper limits on LFV $Z$ decays from LEP and LHC experiments and expected sensitivity of a Tera Z factory as estimated in \cite{Dam:2018rfz} assuming $3\times 10^{12}$ visible $Z$ decays.}
\label{Table:Zdecay_limits}
\end{center}
\end{table}

  In Table~\ref{Table:Zdecay_limits}, we report current limits and future prospects on the three LFVZD modes. As we can see, 
  the old limits obtained by the LEP experiments have been recently superseded by searches performed at the LHC. However, the sensitivity of these latter searches is limited by background events following from $Z\to \tau\tau$ decays to such an extent that at most an improvement by one order of magnitude can be expected after the completion of the future high-luminosity run of the LHC (HL-LHC) (since HL-LHC will collect up to 3000/fb~\cite{Apollinari:2017lan} of integrated luminosity, and the LHC limits in Table~\ref{Table:Zdecay_limits} were obtained with
  $20/{\rm fb}-140/{\rm fb}$). 
  Leptonic colliders instead provide a much more suitable environment to tame such background: for example, the LEP limit on $Z\to\mu e$ obtained in~\cite{Akers:1995gz} was based on a sample of only $4\times 10^6$ $Z$ decays  in a background-free situation, {\it i.e.}, no candidate events were found.\footnote{A major advantage of a leptonic collider is the knowledge of the momenta of the colliding partons, so that the constraint on the invariant mass of the two leptons $m^2_{\ell_i \ell_j} = m_Z^2$ can be precisely implemented up to the beams energy spread, in contrast to the LHC where it is limited by the (large) width of the $Z$.} 
  These processes are therefore an ideal target for future leptonic colliders. 
  In particular, both proposed projects of circular $e^+e^-$ colliders, CEPC~\cite{CEPCStudyGroup:2018rmc,CEPCStudyGroup:2018ghi}
  and FCC-ee~\cite{Abada:2019lih,Abada:2019zxq}, plan to run for several years at a center-of-mass energy around the $Z$ pole, thus acting as a ``Tera Z factory'', {\it i.e.}, collecting $\ord{10^{12}}$ $Z$ decays, about six orders of magnitude more than LEP experiments. 
  The last column of Table~\ref{Table:Zdecay_limits} shows the expected sensitivity of future $e^+e^-$ colliders as estimated in~\cite{Dam:2018rfz} assuming $3\times 10^{12}$ $Z$ decays (corresponding to 150~ab$^{-1}$). 
  As we can see, at least for the $Z\to \tau\ell$ modes, CEPC/FCC-ee could improve on the present LHC (future HL-LHC) bounds up to 4~(3) orders of magnitude, due in particular to an expected excellent momentum resolution ($0.1\%$ at $45\GeV$) of the planned detectors.\footnote{The sensitivity on  BR($Z\to\mu e$) is limited to $\sim 10^{-8}$ by backgrounds from $Z\to\mu\mu$ with one of the muons releasing enough bremsstrahlung energy in the ECAL to be misidentified as an electron~\cite{Dam:2018rfz}. Only in presence of improved
  electron/muon separation methods a sensitivity down to $10^{-10}$ could be achieved.}

Given the outstanding expected sensitivity of future $e^+e^-$ colliders on the $Z\to \ell_i \ell_j$ decays, an obvious question is whether this will be sufficient to test or discover new physics~(NP) scenarios. Indeed, in presence of new physics leading to $Z\to \ell_i\ell_j$, low-energy LFV processes are unavoidably induced by the virtual exchange of the $Z$ itself: $\ell_i \to\ell_j Z^* \to \ell_j f\bar{f}$, where $f$ is a SM quark or lepton. These processes are subject to strong constraints, which then translate into indirect bounds on LFVZD rates~\cite{Nussinov:2000nm,Delepine:2001di,Gutsche:2011bi,Davidson:2012wn}. 
In this work, we plan to reassess the maximal possible LFV effects in $Z$ decays in a model independent way, in view of the improved present and future searches for muon and tau LFV decays.

From the theoretical point of view, the absence of signals in direct searches for the production of new particles at the LHC suggests an energy gap between the electroweak scale and the scale where new physics inducing LFVZD may exist, prompting us to work within the context of an effective field theory~(EFT), {\it i.e.}, introducing a set of higher-dimensional gauge-invariant local operators to be added to the usual SM Lagrangian. 
These operators, built out of SM fields and suppressed by inverse powers of the new physics scale, can parameterise the effects of any kind of NP models as far as the experimentally accessible energies are lower that the actual NP energy scale.
Such an effective theory is known as the Standard Model Effective Field Theory (SMEFT)~\cite{Grzadkowski:2010es,Buchmuller:1985jz} (for a recent review see~\cite{Brivio:2017vri}) and provides the optimal framework for a model-independent analysis. 
In this context, LFV processes could be induced by dimension 6 effective operators, as discussed in detail in~\cite{Raidal:1997hq,Kuno:1999jp,Brignole:2004ah,Ibarra:2004pe,Carpentier:2010ue,Crivellin:2013hpa,Celis:2014asa,Pruna:2014asa,Efrati:2015eaa,Beneke:2015lba,Feruglio:2015gka,Davidson:2016edt,Crivellin:2017rmk,Goto:2015iha,Aebischer:2018iyb,Gonzalez:2021tqc,Cirigliano:2021img,Ardu:2021koz,Etesami:2021hex}. In this article we will employ the SMEFT framework to study low-energy constraints on LFVZD. This type of decays have been also studied within several UV-complete models, such as heavy sterile neutrinos~\cite{Illana:2000ic,Abada:2014cca,Abada:2015zea,DeRomeri:2016gum,Herrero:2018luu,Coy:2018bxr,Hernandez-Tome:2019lkb}, supersymmetry~\cite{Gabbiani:1988pp,Brignole:2004ah}, leptoquarks~\cite{Crivellin:2020mjs}, or in scenarios with extended gauge sectors~\cite{Kuo:1985jt,Langacker:2000ju,Buras:2021btx,Araki:2021vhy}. They have also been previously explored in the context of SMEFT in~\cite{Carpentier:2010ue,Davidson:2012wn,Crivellin:2013hpa,Efrati:2015eaa,Goto:2015iha}. 

The outline of this article is the following. 
In Section~\ref{Sec:setup} we describe the effective field theory setup that has been employed for our analysis. How the SMEFT operators can induce LFVZD is shown in Section~\ref{Sec:LFVZD}. In Section~\ref{Sec:low-energyobs}, we discuss how indirect constraints on $Z\to\ell_i\ell_j$ arise from low-energy LFV observables. The next section contains the results of our analysis, where we assume that the UV physics induces a single dominant operator (Section~\ref{Sec:Single Opr}) or multiple operators that could possibly interfere (Section~\ref{Sec:Double Opr}). We summarise and conclude in Section~\ref{Sec:concl} while a number of useful analytical formulae and results are shown in the Appendix.

\section{Lepton flavour violation in the SMEFT}
\label{Sec:setup}

\begin{table}[t!] 
\centering
\renewcommand{\arraystretch}{1.3}
\setlength{\tabcolsep}{15pt}
\begin{tabular}{cccc} 
 \hline
\multicolumn{2}{c}{4-lepton operators} &
\multicolumn{2}{c}{Dipole operators} \\
\hline
$Q_{\ell\ell }$ & $(\bar L \gamma_\mu L)(\bar L \gamma^\mu L)$ & 
{\color{\tabcolor}$Q_{eW}$} & {\color{\tabcolor}$(\bar L \sigma^{\mu\nu} E) \tau^I \Phi W_{\mu\nu}^I$}  \\
$Q_{ee}$ & $(\bar E \gamma_\mu E)(\bar E \gamma^\mu E)$ &
{\color{\tabcolor}$Q_{eB}$} & {\color{\tabcolor}$(\bar L \sigma^{\mu\nu} E) \Phi B_{\mu\nu}$} \\
$Q_{\ell e}$ & $(\bar L \gamma_\mu L)(\bar E \gamma^\mu E)$ & & \\
\hline
\multicolumn{4}{c}{Lepton-Higgs operators}\\ 
\hline
{\color{\tabcolor}$Q_{\vp \ell }^{(1)}$} & {\color{\tabcolor}$i(\vpj)(\bar L\gamma^\mu L)$} &
{\color{\tabcolor}$Q_{\vp \ell}^{(3)}$} & {\color{\tabcolor}$i(\vpjt)(\bar L \tau^I \gamma^\mu L)$} \\
{\color{\tabcolor}$Q_{\vp e}$} & {\color{\tabcolor}$i(\vpj)(\bar E \gamma^\mu E)$} &
$Q_{e\vp 3}$ & $(\bar L E \Phi)(\Phi^\dag\Phi)$\\
\hline
\multicolumn{4}{c}{2-lepton 2-quark operators} \\
\hline 
$Q_{\ell q}^{(1)}$ & $(\bar L \gamma_\mu L)(\bar Q \gamma^\mu Q)$ & 
  $Q_{\ell u}$ & $(\bar L \gamma_\mu L)(\bar U \gamma^\mu U)$ \\
$Q_{\ell q}^{(3)}$ & $(\bar L \gamma_\mu \tau^I L)(\bar Q \gamma^\mu \tau^I Q)$ & 
$Q_{eu}$ & $(\bar E \gamma_\mu E)(\bar U \gamma^\mu U)$ \\
$Q_{eq}$ & $(\bar E \gamma^\mu E)(\bar Q \gamma_\mu Q)$ &
$Q_{\ell edq}$ & $(\bar L^a E)(\bar D Q^a)$ \\
$Q_{\ell d}$ & $(\bar L \gamma_\mu L)(\bar D \gamma^\mu D)$ & 
$Q_{\ell equ}^{(1)}$ & $(\bar L^a E) \eps_{ab} (\bar Q^b U)$ \\
$Q_{ed}$ & $(\bar E \gamma_\mu E)(\bar D\gamma^\mu D)$ & 
$Q_{\ell equ}^{(3)}$ & $(\bar L^a \sigma_{\mu\nu} E) \eps_{ab} (\bar Q^b \sigma^{\mu\nu} U)$ \\
\hline
\end{tabular}
\caption{ Complete list of the dimension-$6$ operators (invariant under the SM gauge group) which contribute to LFV observables.
Those highlighted in blue generate LFVZD at tree level.
$Q$ and $L$ respectively denote quark and lepton $SU(2)_L$ doublets
($a,b=1,2$ are $SU(2)_L$ indices). $U,\,D$ and $E$ are (up and down) quark and lepton singlets. $\Phi$ represents the Higgs  doublet (and $\vpj\equiv \Phi^\dag (D_\mu \Phi)-(D_\mu \Phi)^\dag \Phi$), while $B_{\mu\nu}$ and $W^I_{\mu\nu}$ are the $U(1)_Y$ and $SU(2)_L$ field strengths, and $\tau^I$ with $I=1,2,3$ are the Pauli matrices. Flavour indices are not shown. 
\label{Tab:dim6}}
\end{table}

Throughout this work, we will assume that the new particles related to the NP scale $\Lambda$ responsible for LFV effects are quite heavy ($\Lambda \gg m_W$) and that there are no other particles in between these scales. 

In such a scenario, it is convenient to work in the SMEFT framework, where the basic idea is to parameterise the low-energy effects of the high-energy theory in terms of higher dimensional operators and the associated Wilson coefficients. 
More specifically, the Lagrangian will consist of that of the SM extended with a tower of higher-dimensional operators suppressed by inverse powers of $\Lambda$:
\be 
\label{Leff} 
\lcal_{\rm SMEFT} = \lcal_{\rm SM} + \frac{1}{\Lambda}
\sum_{a} C_a^{(5)} Q_a^{(5)} + \frac{1}{\Lambda^2} \sum_{a} C_a^{(6)}Q_a^{(6)} + \ocal\left(\frac{1}{\Lambda^3}\right)\,,
\ee
where $\lcal_{\rm SM}$ contains renormalizable operators up to dimension-4, $Q_a^{(n)}$ are the effective operators of dimension-$n$ and the $C_a^{(n)}$ represent the corresponding Wilson coefficients (WCs) which depend on the renormalization scale $\mu$.  
In a given UV-complete model, the Wilson coefficients at the scale $\Lambda$ can be determined by integrating out the heavy particles. 
In our model-independent approach, however, we will consider the $C_a^{(n)}(\Lambda)$ as independent free parameters.
The dominant contributions to our LFV processes are then expected to come from dimension-6 operators,\footnote{There is only one (besides flavour indices) dimension-5 operator, known as the Weinberg operator~\cite{Weinberg:1979sa}, which just induces Majorana neutrino mass terms.} hence we do not consider higher dimension operators in our analysis. Out of the 59 (without counting the combinations of flavour indices) dimension-6 operators~\cite{Buchmuller:1985jz, Grzadkowski:2010es} that can be constructed from SM fields and respect $SU(3)_C\times SU(2)_L\times U(1)_Y$ invariance (and baryon-number conservation), only a subset is relevant to us, namely the operators which contribute to LFV processes at tree-level or at 1-loop level. These are
(i)~4-lepton operators, (ii)~leptonic dipole operators,
(iii)~lepton-Higgs operators, (iv)~2-lepton 2-quark operators. 
A complete list of all 6-dimensional operators that can induce LFV processes as discussed in~\cite{Crivellin:2013hpa} is displayed in~\tref{Tab:dim6}.
In particular,  among these operators, we highlight in blue those that modify the $Z$ boson couplings to leptons and can therefore contribute to $Z \to \ell_i \ell_j$ decays at the tree level, that is,
\begin{itemize}
\item the dipole operators $C^{ij}_{eW}\, Q^{ij}_{eW}$ and $C^{ij}_{eB}  \,Q_{eB}^{ij}$;
\item the lepton-current Higgs-current operators  $C^{(1)\,ij}_{\vp\ell} Q^{(1)\,ij}_{\vp\ell}$, $C^{(3)\,ij}_{\vp\ell} Q^{(3)\,ij}_{\vp\ell}$, and $C^{ij}_{\vp e}\, Q_{\vp e}^{ij}$;
\end{itemize}
where we have explicitly shown the notation for the WCs and flavour indices that we are going to adopt throughout the work\footnote{In the case of 4-fermion operators, there might be redundant flavour combinations, {\it e.g.}, $C_{\ell\ell}^{ijkl}$, $C_{\ell\ell}^{klij}$, $C_{\ell\ell}^{ilkj}$, $C_{\ell\ell}^{kjil}$ giving rise to the same operator (directly or using Fierz identities). We will work in a non-redundant flavour basis and consider only one of these WCs.}.

In order to impose experimental constraints on the coefficients of higher dimensional operators, one needs to evaluate the Renormalisation Group (RG) running from the scale $\Lambda$ to the energy scale relevant for a given experiment. 
As customary, this is done by solving the RG evolution~(RGE) for the higher-dimensional operators, which could not only change the value of a given WC, but also induce mixing between different operators. 
This means generating at low energies operators whose coefficients were set to zero at the scale $\Lambda$.
For a LFV observable, this procedure may consist of up to three steps. 
First, the SMEFT RGE, known at one loop~\cite{Jenkins:2013zja, Jenkins:2013wua, Alonso:2013hga}, will run the WCs from $\Lambda$ to the electroweak scale $\sim m_{Z,W}$.
This running already allows us to extract information relevant for the LFV decays of the $Z$ or Higgs bosons, however it is not enough for energies below the electroweak scale. Therefore, in the second step the SMEFT is matched to the so-called Low-Energy Effective Field Theory (LEFT)~\cite{Jenkins:2017jig,Jenkins:2017dyc}, after integrating out the top quark and the $W^{\pm}$, $Z$ and Higgs bosons.
The final step would then be the QED$\times$QCD running of the LEFT operators to the low-energy scale, $m_\mu$ or $m_\tau$, according to the experimental observable that has to be evaluated. 
We notice that the QED running plays an important role for low-energy LFV observables as it can induce large operator mixing~\cite{Crivellin:2017rmk}.
We implement all these steps for each of the LFV observables entering our numerical analysis with the help of the {\it wilson}~\cite{Aebischer:2018bkb} and {\it flavio}~\cite{Straub:2018kue} packages.

Note that the above lepton-Higgs operators that could be responsible for $Z \to \ell_i \ell_j$ do \emph{not} induce couplings of the physical Higgs particle with leptons, thus they cannot give rise to LFV Higgs decays, $h\to\ell_i \ell_j$. The latter processes can be  induced at the tree level only by the last lepton-Higgs operator of Table~\ref{Tab:dim6}, $Q_{e\vp 3}$, see {\it e.g.}~\cite{Harnik:2012pb,Herrero-Garcia:2016uab}.
Such an operator can be generated by the RG running if at least one of the five operators mentioned above is induced by the UV theory. Nevertheless this effect is proportional to at least one power of a small lepton Yukawa coupling (due to the necessary flip of the chirality of the leptons) and thus their contributions are substantially suppressed~\cite{Jenkins:2013wua,Alonso:2013hga}.\footnote{For example, we find that
$C^{\mu\tau}_{e\vp 3}(m_h) \approx \frac{3 g_1^2}{8 \pi^2}\, y_\tau \log \frac{\Lambda}{m_h} \,C^{(1,3)\,\mu\tau}_{\vp\ell}(\Lambda) \approx 10^{-4}\times C^{(1,3)\,\mu\tau}_{\vp\ell}(\Lambda)$ (for $\Lambda =1$ TeV).}
Similarly, these five operators are not generated by running effects controlled by $C_{e\vp 3}$ (not at one loop). Hence, in the context of the SMEFT, $Z$ and Higgs LFV effects are practically decoupled and we are not going to discuss the latter in this work.

\section{Lepton flavour violating Z decays}
\label{Sec:LFVZD}

The effective interactions involving the $Z$ boson and the SM leptons, including those responsible for LFV effects, are given by the following Lagrangian~\cite{Brignole:2004ah}
\begin{align}
\mathcal L_{\rm eff}^{Z} = &
\Big[ \left(g_{VR}\, \delta_{ij} +\delta g^{ij}_{VR}\right)~ \bar\ell_i\gamma^\mu P_R\ell_j 
\,+\, \left(g_{VL}\, \delta_{ij} +\delta g^{ij}_{VL}\right) ~ \bar\ell_i\gamma^\mu P_L \ell_j \Big] Z_\mu ~+ \nonumber \\
& \Big[\delta g^{ij}_{TR}~ \bar\ell_i\sigma^{\mu\nu} P_R\ell_j\, + \,g^{ij}_{TL}~ \bar\ell_i\sigma^{\mu\nu} P_L\ell_j \Big]  Z_{\mu\nu} ~ + ~ h.c. \,,
\end{align}
where 
\be
g_{VR} =\frac{e\sw}{\cw}
\,,\quad\quad g_{VL} = \frac{e}{\sw\cw} \left(-\frac{1}{2} + \sw^2\right)\,,
\ee
are the SM couplings of $Z$ to right-handed (RH) and left-handed (LH) lepton currents respectively, with $\sw$ ($\cw$) being the sine (cosine) of the weak mixing angle.
New physics effects are encoded in the effective couplings $\delta g_{V/T}$, which at the tree level match the SMEFT operators as follows
\begin{align}
\label{eq:gV}
  &\delta g^{ij}_{VR} = - \frac{ev^2}{2\sw\cw \Lambda^2} \, C^{ij}_{\vp e} \,,
   \quad
    \delta g^{ij}_{VL} = - \frac{ev^2}{2\sw\cw \Lambda^2} \, 
   \Big(C^{(1)\, ij}_{\vp\ell}+C^{(3)\, ij}_{\vp\ell}\Big)\,,
   \\
&    \delta g^{ij}_{TR} = \delta g^{ji\,*}_{TL} = -\frac{v}{\sqrt2 \Lambda^2}\,
    \Big(\sw C_{eB}^{ij} + \cw C_{eW}^{ij}\Big)\,,
    \label{eq:gT}
\end{align}
where the WCs have to be evaluated at the scale $\mu=m_Z$.

The branching ratios of the $Z$ decays into leptons, in particular of the LFV modes, are then given by the following expression~\cite{Brignole:2004ah,Crivellin:2013hpa}
%
%
\begin{equation}
\fontsize{10.5}{6}
\mathrm{BR}\left( {Z \to \ell_i \ell_j } \right) =
\frac{m_Z}{12\pi \Gamma_Z} \Bigg\{
 \left| g_{VR} \delta_{ij} + \delta g^{ij}_{VR} \right|^2 + \left| g_{VL} \delta_{ij} + \delta g^{ij}_{VL} \right|^2  
 + \frac{m_Z^2}{2} \left(\left| \delta g^{ij}_{TR} \right|^2 +
 \left| \delta g^{ij}_{TL} \right|^2\right)
  \Bigg\},
  \label{eq:Zll}
\end{equation}
where $\Gamma_Z = 2.4952(23)$~GeV is the total decay width of the $Z$~boson~\cite{pdg}, and we summed over the two possible combinations of lepton charges, $\ell^\pm_i \ell^\mp_j$. 

As anticipated above, only five SMEFT operators (those highlighted in \tref{Tab:dim6}) can induce the LFV $Z\to \ell_i\ell_j$ decays at the tree level. 
Actually, as we can see from Eqs.~(\ref{eq:gV},\,\ref{eq:gT}), only three independent combinations of the corresponding WCs contribute:
\begin{equation}
\label{eq:comb}
C^{ij}_{\vp e}\,,\quad 
C^{ij}_{\vp \ell} \equiv C^{(1)\,ij}_{\vp \ell}+C^{(3)\,ij}_{\vp \ell}\,,\quad
C_{eZ}^{ij} \equiv \left(\sw C_{eB}^{ij} + \cw C_{eW}^{ij}\right)\,.
\end{equation}

As already mentioned, these WCs are to be evaluated at $\mu=m_Z$. On the other hand, the SMEFT running induces mixing of various operators and therefore the LFVZD will be sensitive to more $C_i(\Lambda)$ beyond those five operators explicitly entering in Eq.~\eqref{eq:Zll}.
In our numerical analysis, we have implemented the full one-loop SMEFT running by means of {\it wilson}~\cite{Aebischer:2018bkb}. 
Nevertheless, given the large number of dimension-6 operators, it is helpful to identify beforehand those WCs that will be more relevant for the LFVZD after the one-loop RG evolution.

For the sake of the following discussion, let us focus on the vectorial couplings in Eq.~\eqref{eq:gV}, and consider only the gauge terms in the one-loop RGE~\cite{Alonso:2013hga}, thus neglecting terms controlled by the Higgs self-coupling and Yukawa couplings.
In that case, the running can induce LFV terms of any of the three Higgs-lepton operators from 24 additional  WCs, although not all of them contributing with the same strength. 
Schematically, we can write it as
\be
\left(\begin{array}{c} \dot \Cone \\ \dot \Ctwo \\ \dot \Cthree \\ \dot\Cfour \end{array}\right) \equiv
16\pi^2 \mu \frac d{d\mu} \left(\begin{array}{c}  \Cone\\  \Ctwo \\  \Cthree\\  \Cfour \end{array}\right) =
\left(\begin{array}{cccc}
\gamma_{11} & \gamma_{12} & 0 & 0 \\
\gamma_{21} & \gamma_{22} & \gamma_{23} & 0 \\
0 & \gamma_{32} & \gamma_{33} & \gamma_{34} \\
0 & 0 & \gamma_{43} & \gamma_{44} 
\end{array}\right)\, 
\left(\begin{array}{c}  \Cone\\  \Ctwo \\  \Cthree \\  \Cfour \end{array}\right) \,,
\label{eq:RGEmat}
\ee
with 
\begin{align}
\Cone \equiv& \Big( C_{\vp \ell}^{(1)}, C_{\vp \ell}^{(3)}, C_{\vp e}\Big)^T,\label{eq:C1}\\
\Ctwo \equiv& \Big(C_{\ell\ell}, C_{\ell e}, C_{ee}, C_{\ell q}^{(1)}, C_{\ell q}^{(3)}, C_{\ell u}, C_{\ell d}, C_{eu}, C_{ed}, C_{qe} \Big)^T, \label{eq:C2}\\ 
\Cthree \equiv & \Big( C_{\vp q}^{(1)}, C_{\vp q}^{(3)}, C_{\vp u}, C_{\vp d}, C_{qq}^{(1)}, C_{qq}^{(3)}, C_{qu}^{(1)}, C_{qd}^{(1)}, C_{uu}, C_{dd}, C_{ud}^{(1)} \Big)^T,\label{eq:C3}\\
\Cfour \equiv & \Big(C_{qu}^{(8)}, C_{qd}^{(8)}, C_{ud}^{(8)} \Big)^T,\label{eq:C4}
\end{align}
and the matrices $\gamma_{ij}$ encoding the anomalous dimensions, whose explicit form can be obtained from Ref.~\cite{Alonso:2013hga}.
Despite the fact that the RGE will mix these 27 operators, from the structure of Eq.~\eqref{eq:RGEmat} it is clear that the contribution from the fourteen WCs in $\Cthree$ and $\Cfour$ will be negligible for the LFVZD, as they do not generate directly any Higgs-lepton operator.
The vectorial contributions to the LFV Z decays will then come mainly from three WCs in Eq.~\eqref{eq:C1} and one-loop suppressed contribution from ten WCs in Eq.~\eqref{eq:C2}.

A similar classification can also be done for the dipole operators $C_{eW}$ and $C_{eB}$.
In this case, we can identify a set of 10 operators forming a closed RGE system, including themselves and other operators involving gauge and Higgs fields, not shown in~\tref{Tab:dim6}. 
As we will show in Section~\ref{Sec:Single Opr}, however, dipole operators are so constrained by low-energy LFV observables that result completely irrelevant for the sake of LFVZD, and we therefore refrain from discussing further their RGE effects.

\section{Indirect constraints from low-energy LFV observables}
\label{Sec:low-energyobs}
\begin{figure}[t!]
\begin{center}
\includegraphics[width=.95\textwidth]{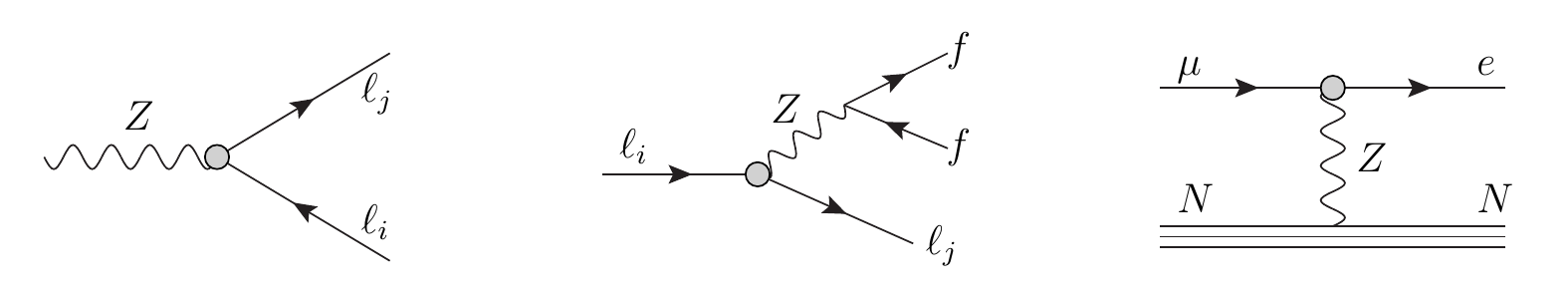}
\caption{
Examples of tree-level diagrams where the same LFV vertex generating $Z\to\ell_i\ell_j$ induces low-energy processes such as $\ell_i\to\ell_j ff$, with $f=\ell,q$, or $\mu$-$e$ conversion in nuclei.}
\label{diagsLFV}
\end{center}
\end{figure}
The operators contributing to the LFV Z decays also contribute to low-energy LFV observables, as shown in Fig.~\ref{diagsLFV}.
Among the most relevant observables, we find the $\mu$ to $e$ conversion in nuclei, the radiative decays $\ell_i\to\ell_j\gamma$, the leptonic decays $\ell_i\to\ell_j\ell_k\bar\ell_m$,  and the semileptonic decays $\tau\to\ell M$, with $M$ being a pseudoscalar or vectorial meson.
All these processes are severely constrained experimentally, see \tref{Tab:LFVlimits}, and thus set indirect constraints on the WCs we are considering and on the maximum allowed rate for the LFV $Z$ decays.\footnote{Here we do not consider processes that violate individual flavour numbers by two units, $\Delta L_i =2$, such as muonium-antimuonium oscillations (for a recent discussion in the context of SMEFT see~\cite{Conlin:2020veq}), since the operators we are interested in do not induce them at leading order and, as we will see, they are so tightly constrained that their possible one-loop contributions to $\Delta L_i =2$ are extremely suppressed.} 

\begin{table}[t]
\begin{center}
{\small
\renewcommand{\arraystretch}{1.3}
\begin{tabular}{lclrl}
\hline
LFV obs. & \multicolumn{2}{l}{Present bounds  $(90\%\,\text{CL})$} & \multicolumn{2}{l}{Expected future limits} \\
\hline
BR$(\mu\to e\gamma)$ &  $4.2\times10^{-13}$ &MEG (2016)~\cite{TheMEG:2016wtm} & $6\times 10^{-14} $& MEG-II~\cite{Baldini:2018nnn}\\
BR$(\mu\to eee)$ &   $1.0\times10^{-12}$ & SINDRUM (1988)~\cite{Bellgardt:1987du} & $10^{-16}$ &Mu3e~\cite{Blondel:2013ia} \\
CR$(\mu\to e,{\rm Au})$ & $7.0\times10^{-13}$ &SINDRUM II (2006)~\cite{Bertl:2006up}&  --\hspace{.7cm}  & -- \\
CR$(\mu\to e,{\rm Al})$ & -- & -- & $6\times10^{-17}$ &COMET/Mu2e~\cite{Kuno:2013mha,Bartoszek:2014mya} \\
\hline
BR$(\tau\to e\gamma)$ & $3.3\times10^{-8}$ &BaBar (2010)~\cite{Aubert:2009ag} & $ 3\times 10^{-9\phantom{0}}$ &Belle-II~\cite{Kou:2018nap}\\
BR$(\tau\to eee)$ & $2.7\times10^{-8}$ &Belle (2010)~\cite{Hayasaka:2010np} & $5\times 10^{-10}$ &Belle-II~\cite{Kou:2018nap}\\
BR$(\tau\to e\mu\mu)$ & $2.7\times10^{-8}$ &Belle (2010)~\cite{Hayasaka:2010np} & $5\times 10^{-10}$ &Belle-II~\cite{Kou:2018nap}\\
BR$(\tau\to \pi e)$ & $8.0\times10^{-8}$  &  Belle (2007)~\cite{Miyazaki:2007jp} & $4\times 10^{-10}$ &Belle-II~\cite{Kou:2018nap}\\
BR$(\tau\to \rho e)$ & $1.8\times10^{-8}$ &Belle (2011)~\cite{Miyazaki:2011xe} & $3\times 10^{-10}$ &Belle-II~\cite{Kou:2018nap}\\
\hline
BR$(\tau\to \mu\gamma)$ & $4.2\times10^{-8}$ &Belle (2021)~\cite{Belle:2021ysv} & $ 10^{-9\phantom{0}}$ &Belle-II~\cite{Kou:2018nap}\\
BR$(\tau\to \mu\mu\mu)$ & $2.1\times10^{-8}$ &Belle (2010)~\cite{Hayasaka:2010np} & $4\times 10^{-10}$ &Belle-II~\cite{Kou:2018nap}\\
BR$(\tau\to \mu ee)$ & $1.8\times10^{-8}$ &Belle (2010)~\cite{Hayasaka:2010np} & $3\times 10^{-10}$ &Belle-II~\cite{Kou:2018nap}\\
BR$(\tau\to \pi\mu)$ &  $1.1\times10^{-7}$ & Babar (2006)~\cite{Aubert:2006cz} & $5\times 10^{-10}$ &Belle-II~\cite{Kou:2018nap}\\
BR$(\tau\to \rho\mu)$ & $1.2\times10^{-8}$ &Belle (2011)~\cite{Miyazaki:2011xe} & $2\times 10^{-10}$ &Belle-II~\cite{Kou:2018nap}\\
\hline
\end{tabular}
\caption{Present upper bounds and future expected sensitivities for the set of low-energy LFV transitions relevant for our analysis.
\label{Tab:LFVlimits}}
}
\end{center}
\end{table}

As we will be dealing with low-energy observables, we need to use the full QFT machinery described in \sref{Sec:setup}, {\it i.e.}, we need to run the WCs at $\mu=\Lambda$ down to the electroweak scale using the SMEFT running, then match them to the LEFT operators and finally use the LEFT running down to $\mu=m_\tau$ or $m_\mu$. 
A complete LEFT basis, the one-loop RGE and the tree-level matching to SMEFT operators can be found in Refs.~\cite{Jenkins:2017jig,Jenkins:2017dyc}.
In our case, the most relevant LEFT operators will consist of four fermions with at least one spinor combination consisting of  two different lepton flavours, and the photon dipole operator. 
Schematically, the former will take the form
\be
\label{LEFT4fermion}
\ocal_{AXY}= \big(\overline{\ell_i}\,\Gamma_{A}P_X \ell_j\big)\big(\overline{f_{\alpha}}\,\Gamma_{A}P_Y f_{\beta}\big)\,, 
\ee
where $f_{\alpha,\beta}$ refers to any light quark or lepton, $P_{X,Y}$ to left or right projectors and $\Gamma_A$ to the scalar, vector or tensor operators, {\it i.e.}, $\Gamma_S=\mathbf 1, \Gamma_V = \gamma_\mu$ or $\Gamma_T = \sigma_{\mu\nu}$. 
On the other hand, the relevant photon dipole operator is given by
\be
\label{LEFTdipole}
\ocal_{\gamma} = \overline{\ell_i}\, \sigma^{\mu\nu} P_R \ell_j\, F_{\mu\nu}~ (+ h.c.)\,,
\ee
which directly matches, at tree-level, to the orthogonal combination of $C_{eZ}$, cf.~\eref{eq:comb}, {\it i.e.},
\be\label{SMEFTphotondipole}
C_{\gamma}^{ij} = \frac v{\sqrt2 \Lambda^2} \left( \cw C_{eB}^{ij} - \sw C_{eW}^{ij}\right) \equiv \frac v{\sqrt2 \Lambda^2} C_{e\gamma}^{ij}\,.
\ee
The case of the dipole has been extensively studied also beyond leading order, showing the relevance of performing the matching at one-loop~\cite{Pruna:2014asa} and including the {\it leading} two-loop anomalous dimensions~\cite{Crivellin:2017rmk}.
Nevertheless, we will see that the naive analysis including only the tree-level matching and the one-loop anomalous dimensions already suggests that the role of the dipoles will be suppressed in the LFVZD due to the strong constrains from the radiative decays.
Therefore, we will not include any of those higher order contributions in our numerical analysis.

Once again, all this procedure is included in our numerical analysis using {\it wilson} and {\it flavio}.
Nevertheless, and with the aim of improving our understanding of the numerical results, we give in \aref{App:lowLFV} the analytical expressions for these observables in terms of the $C_i(\Lambda)$ after tree-level matching, but neglecting the RGE effects. 
We checked that these formulas provide a good approximation for the leading contributions, {\it e.g.} the contributions of $C_{\vp e}$ or $C_{\ell u}$ to $\tau\to\rho\ell$, and will describe the overall behaviour of most of our numerical results. 
Note however that they neglect RGE-induced mixing that might be important in some cases, such as the role of $C_{\ell\ell}$ in $\tau\to\rho\ell$, which is a consequence of the SMEFT running in Eq.~\eqref{eq:RGEmat} and the equivalent LEFT one. 
We will discuss the relevance of these latter contributions while presenting our numerical analysis.

Finally, it is worth mentioning that the operators shown in Table \ref{Tab:dim6} can also be probed at high-energy colliders, in particular the 2-lepton 2-quark operators that we are going to consider in the next section. Nevertheless, current LHC sensitivities for the operators we are interested in are not competitive with those at low-energy experiments~\cite{Angelescu:2020uug}. 

\section{Results and discussion}
\label{Sec:Results}

In \sref{Sec:LFVZD} we have seen that there are five SMEFT operators, in three independent combinations displayed in \eref{eq:comb}, that directly generate LFV $Z$ decays at the tree level and therefore they will give the dominant contributions to our observables.
For this reason, it is interesting to analyse their effect when only one of these operators is present, assuming  other operators are absent at that time, and to compare the potential of the LFVZD to that of other low-energy LFV observables in the search for new physics. 
Nevertheless, it is reasonable to expect that a given UV theory will generate several SMEFT operators at the same time.
In such a case, possible interference between different operators could distort the results obtained considering each operator individually. 
In order to illustrate this idea, we will consider two non-zero operators at a time, paying special attention to possible flat directions, {\it i.e.}, fine cancellations among contributions stemming from different operators. 
We will see how other operators beyond those in~\eref{eq:comb} could play an important role in this case, and we will assess whether they could suppress some of the low-energy LFV decays, thus allowing larger LFV $Z$ decay rates.\footnote{In what follows, we assume real WCs for simplicity. In fact, the total decay rates for the LFV $Z$ decays can not be affected by the phase of the WCs, and, even in case of discovery, we expect that the number of LFVZD events would be insufficient to perform an analysis in search of CP-violating effects.} 

\subsection{Single operator dominance}
\label{Sec:Single Opr}

\begin{table}[t!]
\begin{center}
\renewcommand{\arraystretch}{1.3}
\begin{tabular}{clcc}
\hline
~~~~Observable~~~~ & \phantom{\big(}Operator &  ~~Indirect Limit on LFVZD~~ &  Strongest constraint   \\
\hline
\multirow{4}{*}{BR($Z\to \mu e$)} 
& $\big(Q^{(1)}_{\vp\ell}+Q^{(3)}_{\vp\ell}\big)^{e\mu}$  & $3.7\times 10^{-13} $  &  $\mu  \to e,\,\text{Au}$   \\
& \phantom{\big(}$Q^{e\mu}_{\vp e}$  & $9.4\times 10^{-15} $  &  $\mu  \to e,\,\text{Au}$ \\
& \phantom{\big(}$Q^{e\mu}_{eB}$  & $1.4\times 10^{-23} $  &  $\mu  \to e\gamma$ \\
& \phantom{\big(}$Q^{e\mu}_{eW}$  & $1.6\times 10^{-22} $  &  $\mu  \to e\gamma$ \\
\hline
\multirow{4}{*}{BR($Z\to \tau e$)} & 
$\big(Q^{(1)}_{\vp\ell}+Q^{(3)}_{\vp\ell}\big)^{e\tau}$ & $6.3\times 10^{-8} $  &  $\tau \to \rho \,e$   \\
& \phantom{\big(}$Q^{e\tau}_{\vp e}$ & $ 6.3\times 10^{-8} $  &  $\tau \to \rho \, e$   \\
& \phantom{\big(}$Q^{e\tau}_{eB}$ & $1.2\times 10^{-15} $  &  $\tau \to  e\gamma$   \\
& \phantom{\big(}$Q^{e\tau}_{eW}$ & $1.3\times 10^{-14} $  &  $\tau \to  e\gamma$   \\
\hline
\multirow{4}{*}{BR($Z\to \tau \mu$)} &
$\big(Q^{(1)}_{\vp\ell}+Q^{(3)}_{\vp\ell}\big)^{\mu\tau}$& $4.3\times 10^{-8} $  &  $\tau \to \rho \,\mu$   \\
& \phantom{\big(}$Q^{\mu\tau}_{\vp e}$ & $4.3\times 10^{-8} $  &  $\tau \to \rho \, \mu$   \\
& \phantom{\big(}$Q^{\mu\tau}_{eB}$ & $1.5\times 10^{-15} $  &  $\tau \to \mu\gamma$   \\
& \phantom{\big(}$Q^{\mu\tau}_{eW}$ & $1.7\times 10^{-14} $  &  $\tau \to \mu\gamma$   \\
\hline
\end{tabular}
\caption{Indirect upper limits on $\text{BR}(Z\to \ell_i \ell_j)$ considering a single operator at the scale $\mu=m_Z$.
The last column shows which low-energy observable gives the strongest constraint.
These indirect limits are to be compared with the future expected bounds at a Tera Z factory shown in Table~\ref{Table:Zdecay_limits}, {\it i.e.} BR$(Z\to \mu e)<10^{-8}-10^{-10}$ and BR$(Z\to\tau\ell)<10^{-9}$. 
}\label{Table:Single operators bound}
\end{center}
\end{table}

We want to analyse the effect of the five relevant SMEFT operators for tree-level LFVZD, {\it i.e.}, the two dipoles and three Higgs-lepton operators, assuming that only one of them is present at a time, which would be approximately the case if the underlying UV dynamics mostly matches to a single operator while inducing others at a substantially suppressed level.
This hypothesis, however, needs to be defined at a given scale $\mu$. 

We start considering a single non-zero WC at $\mu=m_Z$, as this is the relevant scale for the $Z$ decays.
\tref{Table:Single operators bound} shows the results we found for the maximum allowed LFVZD rates after requiring that all the the low-energy LFV processes that are induced by the same operators lie below their current experimental limits shown in \tref{Tab:LFVlimits}. 
Notice that, in this case, the RGE effects do not affect the LFVZD, which can be directly computed by means of \eref{eq:Zll}. 
This is not true for the low-energy observables, for which the proper matching and RGE  to $\mu=m_\tau$ or $m_\mu$ need to be taken into account, as discussed in \sref{Sec:setup}.

From \tref{Table:Single operators bound} we see that dipole operators are extremely constrained by the radiative decays $\ell_i\to\ell_j\gamma$, which translates to LFVZD rates beyond current and future experimental sensitivities. 
At this point, one could be tempted to tune $C_{eB}$ and $C_{eW}$ in such a way that the photon dipole in \eref{SMEFTphotondipole} vanishes, avoiding the bounds from $\ell_i\to\ell_j\gamma$ and maximizing LFVZD via the $Z$ dipole.
Nevertheless, this choice would have several drawbacks. 
Firstly, it seems very unlikely to have a UV model that leads to a vanishing $C_{e\gamma}(m_Z)$. 
Notice that even if the UV model generated only $C_{eZ}$ at the NP scale $\mu=\Lambda$, the RGE would induce a non-zero photon dipole at $\mu=m_Z$. This means that a huge fine-tuning between $C_{e\gamma}$ and the radiative effects would be needed to have $C_{e\gamma}(m_Z)=0$.
Secondly, a vanishing photon dipole would only suppress the tree-level contributions to $\ell_i\to\ell_j\gamma$, however higher order terms would still be important~\cite{Crivellin:2013hpa,Pruna:2014asa,Crivellin:2017rmk}.
Although not included in \tref{Table:Single operators bound}, we have estimated the size of these higher order effects following~\cite{Pruna:2014asa} and found that the radiative decays would still impose strong bounds even in the extreme case of vanishing $C_{e\gamma}(m_Z)$, setting indirect limits on dipole mediated LFVZD beyond future sensitivities.

On the other hand, Higgs-lepton operators, which do not generate $\ell_i\to\ell_j\gamma$ at the tree-level,\footnote{As in the case of $C_{eZ}$, these operators induce $\ell_i\to\ell_j\gamma$ at higher order and may still be constrained by these processes. Nevertheless, we checked that these bounds are weaker than those coming from tree-level mediated processes such as $\mu$-$e$ conversion in nuclei or $\tau\to\rho\ell$.} are less constrained and larger LFVZD are allowed. 
In the $\mu$-$e$ sector, the strongest current bounds are imposed by $\mu$-$e$ conversion in nuclei. This translates into an indirect bound of BR($Z\to\mu e)\lesssim 10^{-13}$, which unfortunately is still beyond the reach of future experiments, see \tref{Table:Zdecay_limits}.
The results for the tau sector are however more optimistic. 
In this case, they are currently mostly constrained by $\tau\to\rho \ell$ decays, which imposes indirect limits of the order BR($Z\to\tau \ell) \lesssim  10^{-8}-10^{-7}$.
While still below the reach of current LEP/LHC bounds (as well as the expected HL-LHC sensitivity), these decay rates could be probed at a future Tera~Z factory.

\begin{figure}[t!]
\begin{center}
\includegraphics[width=.9\textwidth]{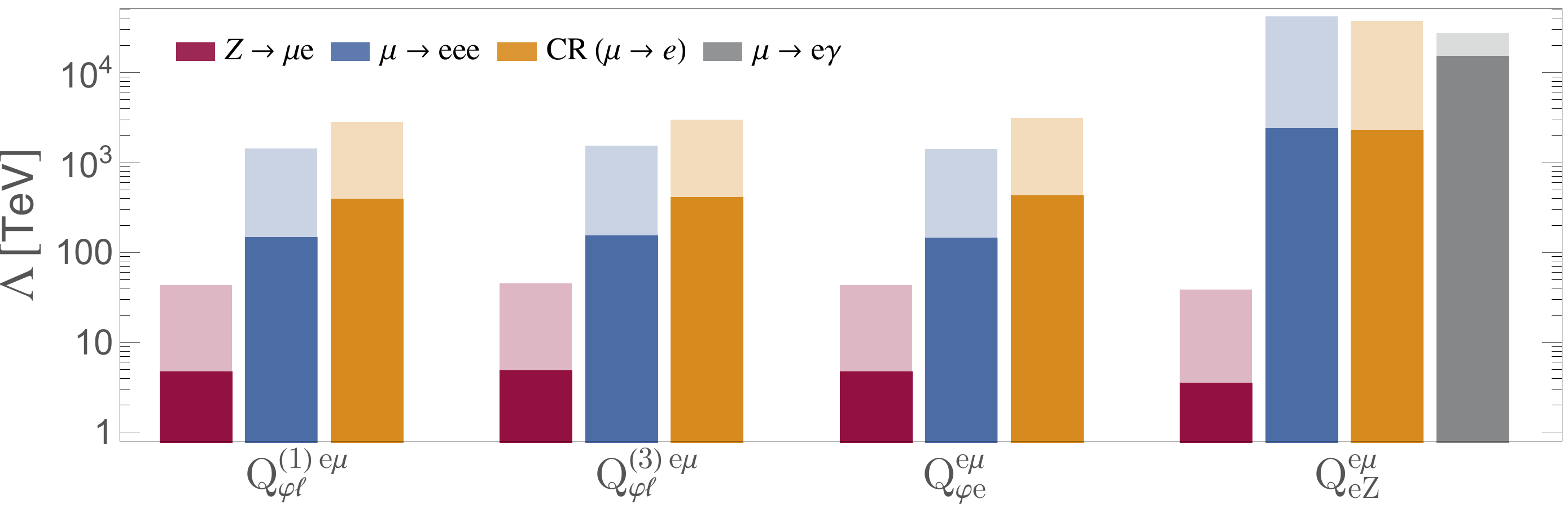}\\
\includegraphics[width=.9\textwidth]{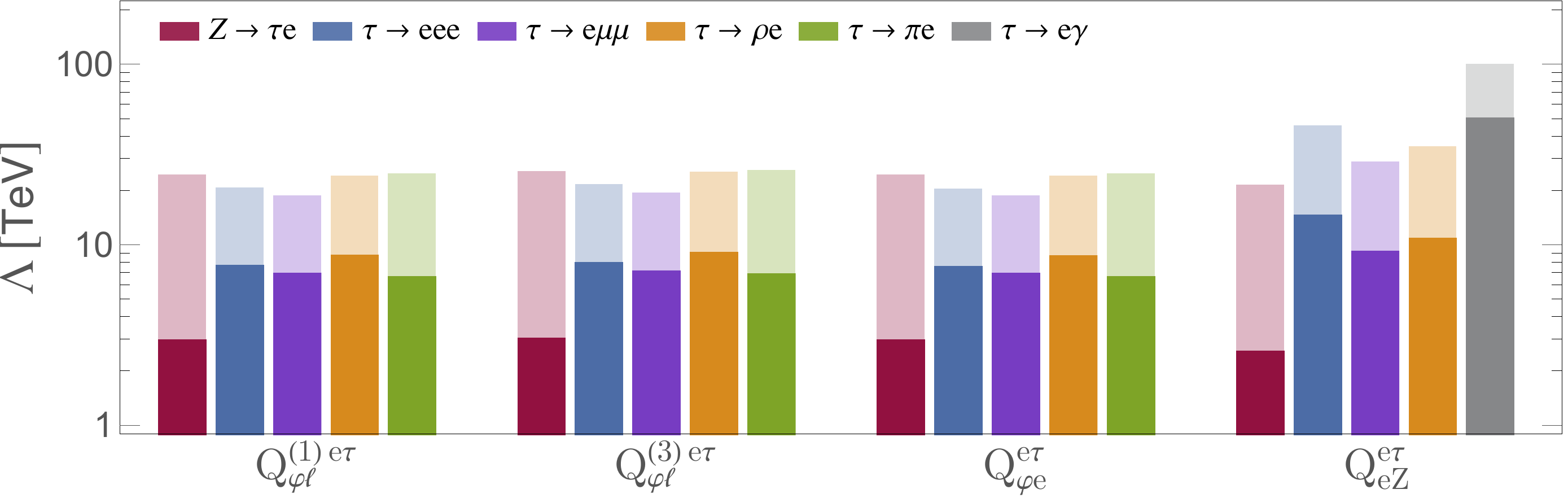}\\
\hspace{.01cm}
\includegraphics[width=.9\textwidth]{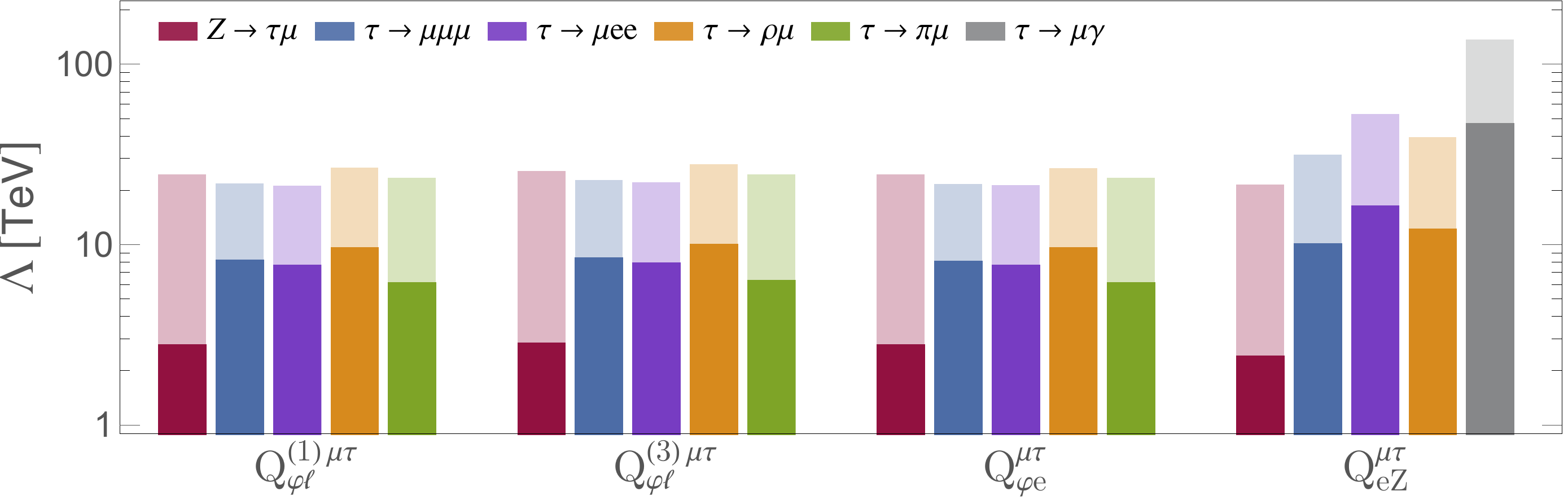}
\caption{
Values of the NP scale $\Lambda$ that are accessible by each of the LFV observables with current bounds (solid bars) and future sensitivities (lighter bars). 
We assume that $C(\Lambda)\leq1$ for each operator at a time, while the others vanish at $\mu=\Lambda$.
\label{Fig:Single operators dominance}}
\end{center}
\end{figure}

Next, we consider as input the WCs at the NP scale $\mu=\Lambda$, as it would be the scale at which we could integrate out the new heavy fields related to the UV dynamics and generate the SMEFT operators.
In order to compare better the sensitivity reach of future high- and low-energy LFV experiments to such operators, we focus our discussion on what would be the largest NP scale that we could probe in each case. 
Under our hypothesis of switching on a single operator at a time, the LFV observables under consideration would scale as $C(\Lambda)^2/\Lambda^4$, up to some $\mathcal O (\log\mu/\Lambda)$ corrections from the RGE.
This means that, for a given experimental upper limit on an observable, the maximum NP scale that we could be probing  corresponds to the maximum value for the WC at that scale.
Notice that higher NP scales would require non-perturbative WCs, while smaller scales would always be allowed provided that the WC is small enough. 
We show in \fref{Fig:Single operators dominance} the sensitivities for $\Lambda$ for our most relevant operators and from different observables, where we have assumed that $C(\Lambda)\leq1$ from perturbativity arguments.
In this case, and opposite to \tref{Table:Single operators bound}, we choose the $Z$ dipole operator as input, which implicitly assumes $C_{e\gamma}(\Lambda)=0$, since this hypothesis is still challenging but more plausible at $\mu=\Lambda$.
We also show $Q^{(1)}_{\vp\ell}$ and $Q^{(3)}_{\vp\ell}$ separately as they have different RGE. Nevertheless, the differences are numerically small and difficult to appreciate in the Figure.

From \fref{Fig:Single operators dominance} we can see that current sensitivities (solid bars) are always worse in the case of the LFVZD than from low-energy observables, in agreement with our findings in \tref{Table:Single operators bound}, and especially in the case of the dipoles. 
Despite we chose to switch on only the $Z$ dipole and not the photon one at $\mu=\Lambda$, the RGE generate a photon dipole at low energies, providing a better sensitivity to NP from low-energy observables even in this extreme case. 
Unfortunately, the situation will be similar for future experiments, therefore we can conclude that LFVZD will not be competitive probing the dipole operators and ignore them for the rest of the analysis. 

On the other hand, the Higgs-lepton operators are again more promising, in particular in the $\tau$-$\ell$ sectors. 
We can clearly see how the huge improvement in sensitivities at the Tera Z factories will boost the potential of the $Z\to\tau\ell$ decays, reaching values that are competitive with low-energy observables. 
This result is remarkable and very promising, since the future limits for LFV $\tau$ decays shown in \tref{Tab:LFVlimits} are based on the most optimistic assumption that Belle-II searches will be background-free, which does not necessarily need to be the case. Therefore, being able to access the same NP scale with an independent high-energy observable would be of great value. 
This will not be the case, however, in the $\mu$-$e$ sector, where the future reach on $Z\to \mu e$ lies below the current bounds from low-energy processes, even considering the most optimistic sensitivity BR$(Z\to \mu e)\sim 10^{-10}$ as we did in \fref{Fig:Single operators dominance}. 
Indeed, the outstanding future reach of $\mu\to eee$ and $\mu\to e$ conversion experiments will probe scales almost two orders of magnitude above those accessible at Tera Z factories through $Z\to \mu e$.

In summary, we have seen that if a single operator dominates, future Tera Z factories searching for LFVZD could probe NP at the same level as low-energy experiments, in particular for the $\tau$-$\ell$ sector of the Higgs-lepton operators. 
Notice that so far we have only considered the leading order operators for LFVZD, however we have seen that other operators can also be relevant due to RGE effects, in particular those of \eref{eq:C2}.
Under this single operator dominance hypothesis, we can expect that these new operators lead to small LFVZD rates, as they are constrained by processes that they do generate at tree level.
Nevertheless, they could still induce new interesting effects when combining several operators at a time, as we are going to explore in the following.

\subsection{Interference of multiple operators}\label{Sec:Double Opr}
We now move to consider possible interference effects arising when multiple operators are induced by the UV dynamics at the scale $\Lambda$. As argued above, in the case of dipole operators constraints from LFV radiative decays are so difficult to overcome that we cannot expect these operators to be a substantial source of LFVZD. In the following discussion, we therefore focus on the three remaining Higgs-lepton operators. 

As we have seen, the combination  
$C^{(1)\,ij}_{\vp \ell}(m_Z)+C^{(3)\,ij}_{\vp \ell}(m_Z)$ contributes to $Z\to \ell_i \ell_j$. Furthermore, we notice that, after decoupling the $Z$ boson, the very same combination, at the same scale, also matches to the LEFT 4-fermion operators relevant for low-energy LFV processes, as can be seen by inspecting the formulae in the Appendix~\ref{App:lowLFV}. As a consequence, even in presence of a cancellation between $C^{(1)\,ij}_{\vp \ell}(m_Z)$ and $C^{(3)\,ij}_{\vp \ell}(m_Z)$, this would similarly affect both $Z\to \ell_i \ell_j$ and the indirect constraints, thus without modifying their relative importance shown in \fref{Fig:Single operators dominance}.
For the same reason, although very subdominant differences stemming from the RGEs of the two WCs, the effects of $C^{(1)}_{\vp \ell}$ and $C^{(3)}_{\vp \ell}$ are practically identical for both LFVZD and low-energy processes. Hence, in the following, we will only show results for $C^{(1)}_{\vp \ell}$.
Notice also that the RH operator $C_{\vp e}$ cannot interfere with LH Higgs-lepton operators neither for LFVZD, see \eref{eq:Zll}, nor for low-energy LFV processes, cf.~Appendix~\ref{App:lowLFV}. Therefore, if the UV physics mostly induces Higgs-lepton operators no cancellations are possible and the indirect limits reported in Table~\ref{Table:Single operators bound} are still valid if $C^{(1)\,ij}_{\vp \ell}(m_Z)$, $C^{(3)\,ij}_{\vp \ell}(m_Z)$ and $C^{ij}_{\vp e}(m_Z)$ are all non-vanishing. 
In order to study the possibility of non-trivial cancellations, we then have to consider simultaneous presence at the scale $\Lambda$ of a non-zero coefficient of a Higgs-lepton operator (we start with $C^{(1)}_{\vp \ell}$) and one or more 4-fermion SMEFT operators, such as 4-lepton or 2-lepton 2-quark operators (cf.~Table~\ref{Tab:dim6}) that are relevant for the low-energy constraints.

\begin{figure}[t!]
\centering
\includegraphics[width=0.9\textwidth]{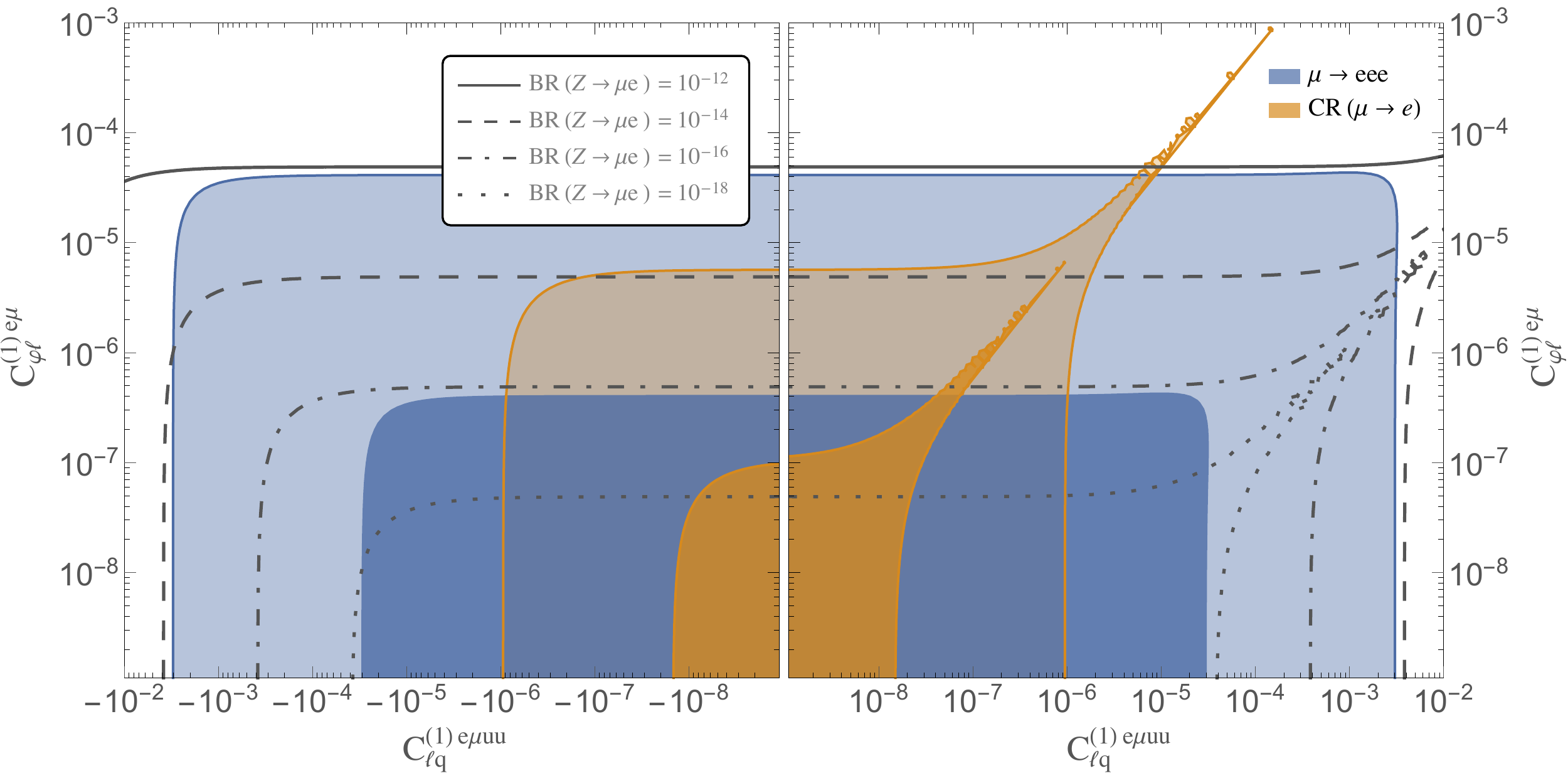}
\includegraphics[width=0.9\textwidth]{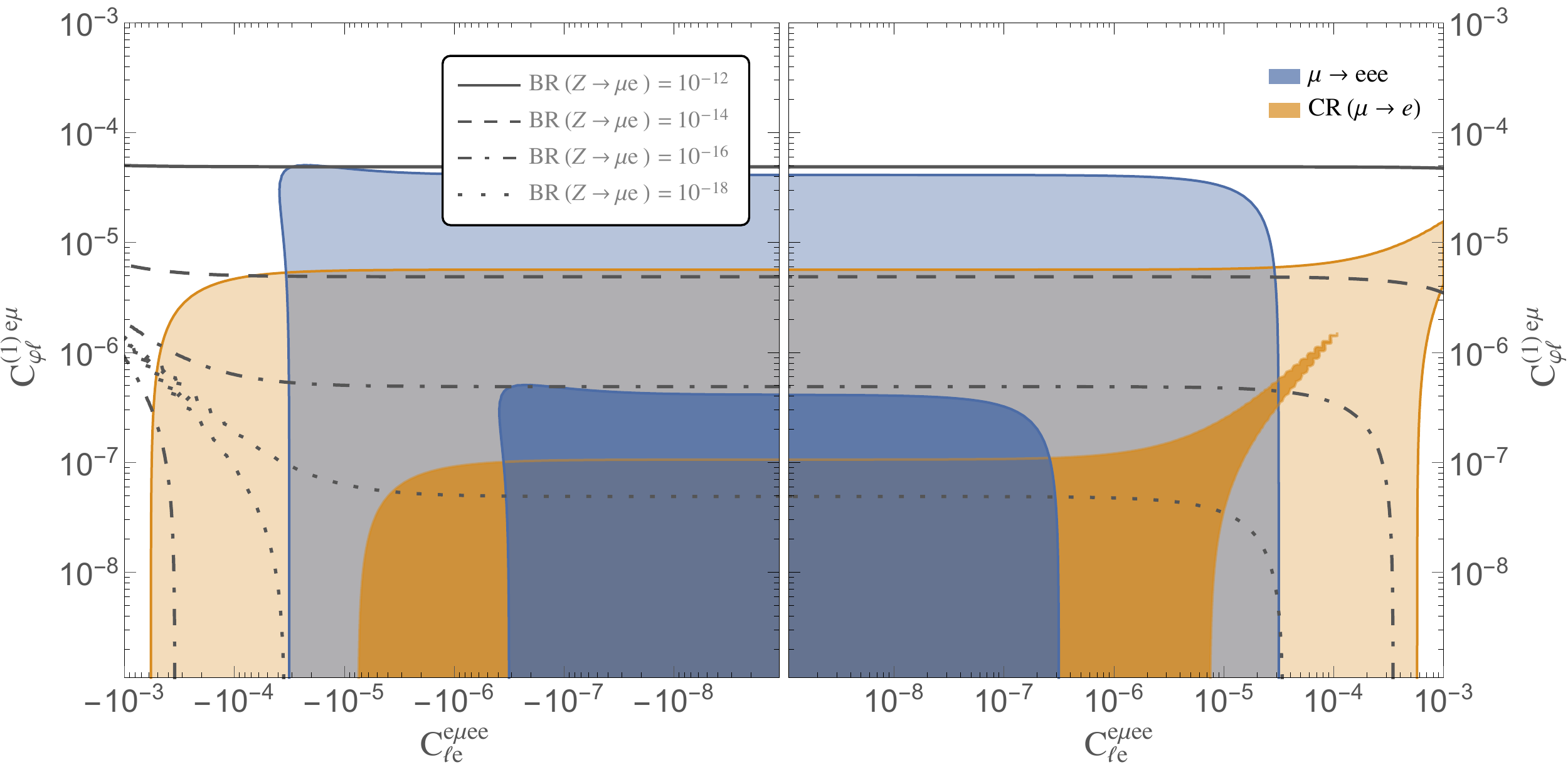}
\caption{Contours of BR($Z\to \mu e$) as a function of  $C^{(1)\,e\mu}_{\vp \ell}(\Lambda)$ and $C^{(1)\,e\mu uu}_{\ell q}(\Lambda)$  (top panel), $C^{(1)\,e\mu}_{\vp \ell}(\Lambda)$ and  $C^{e\mu ee}_{\ell e}(\Lambda)$ (bottom panel). The cutoff scale is set to $\Lambda = 1$~TeV. The lighter coloured regions are currently {\it allowed} by $\mu\to eee$ (blue) and $\mu\to e$ conversion (orange), the corresponding darker regions show the future expected sensitivities.
\label{fig:results-mue}
}
\end{figure}

\paragraph{$\boldsymbol{\mu-e}$ sector.} 
In \fref{fig:results-mue}, we plot contours of BR($Z\to \mu e$) on the plane $C^{(1)\,e\mu}_{\vp \ell}(\Lambda)$ versus the WC of the 2-lepton 2-quark operator $C^{(1)\,e\mu uu}_{\ell q}(\Lambda)$  (top panel) and the 4-lepton operator $C^{e\mu ee}_{\ell e}(\Lambda)$ (bottom panel).
We choose $\Lambda = 1$~TeV for this and the following Figures, nevertheless our results are qualitatively the same for other scales, following the scaling $C_i(\Lambda)/\Lambda^2$ with small logarithmic modifications from the RGE.
The lighter coloured regions are currently \emph{allowed} by the $\mu\to eee$ (blue) and $\mu\to e$ conversion (orange) constraints, while the corresponding darker regions show how the allowed parameter space will shrink given the future expected sensitivities reported in Table~\ref{Tab:LFVlimits}. 

From the examples in these plots, we can see that a value of the 4-fermion WC at the scale $\Lambda = 1$~TeV approximately three orders of magnitude larger than the value of $C^{(1)\,e\mu}_{\vp \ell}(\Lambda)$ could in principle conspire to give $C^{(1)\,e\mu}_{\vp \ell}(m_Z) \approx 0$, hence suppressing the BR$(Z\to \mu e$), due to a cancellation between $C^{(1)\,e\mu}_{\vp \ell}(\Lambda)$ and its RGE running from $\Lambda$ to $m_Z$, cf.~\sref{Sec:LFVZD}. However, we can also see that such cancellations would occur for values of the parameters that are already excluded by the combination of the bounds from $\mu\to eee$ and $\mu\to e$ conversion.

The plots of \fref{fig:results-mue} also show how difficult it is to tune the parameters in such a way that LFVZD effects are enhanced relative to the muon LFV processes. From the top panel
we can see that there is a flat direction cancelling the conversion rate of $\mu \to e$ in nuclei for  
$C^{(1)\,e\mu uu}_{\ell q}(\Lambda) \approx 0.2 \times C^{(1)\,e\mu}_{\vp \ell}(\Lambda)$,
as a consequence of the coherent nature of the conversion process and the consequent interference of the various contributions shown in~\eref{eq:mueconv}.
This numerical result is well reproduced by the analytical approximate formulae collected in the \aref{App:mueconversion}. From these expressions it is easy to check that the cancellation requires
\be
\label{eq:flatD}
C^{(1)\,e\mu uu}_{\ell q}(\Lambda) \approx \frac{V^{(n)}-(1-4 \sw^2)V^{(p)}}{3(V^{(n)}+V^{(p)})} \times C^{(1)\,e\mu}_{\vp \ell}(\Lambda)\,,
\ee
if only these vector operators are involved as in our example. This also shows that a tuning of the parameters to obtain CR$(\mu\to e,\,N)\approx 0$ for a certain nucleus would \emph{not} lead to an exact cancellation of the conversion rate in another nucleus, for which the overlap integrals $V^{(n)}$ and $V^{(p)}$ have different numerical values~\cite{Kitano:2002mt}.\footnote{Notice that indeed the displayed current and future bounds follow from muon conversion in different nuclei, respectively Au and Al, cf.~\tref{Tab:LFVlimits}. However, the quantity appearing in the right-hand side of \eref{eq:flatD} is numerically similar in the two cases (0.19 for Au, 0.17 for Al), such that the slight difference in the flat direction is difficult to appreciate in a logarithmic plot as those in \fref{fig:results-mue}.}
Moreover, $\mu\to eee$ is clearly unaffected by such a tuning as shown by the figure. In the displayed example, this implies an indirect bound $\text{BR}(Z\to \mu e)< 10^{-12}$ even along the flat direction, which, although being an order of magnitude better with respect to the single operator analysis in \tref{Table:Single operators bound}, still lies beyond the expected reach at the Tera Z factory.

The bottom panel of \fref{fig:results-mue} shows what happens if the dominant operator generated at high energies alongside $C^{(1)\,e\mu}_{\vp \ell}$ is a 4-lepton operator. An exact cancellation of $\mu\to e$ conversion in nuclei is still possible, albeit it is due to radiative effects in this case. Indeed, the RGE running below the electroweak scale induces 2-lepton 2-quark operators through loops involving the 4-lepton operator, with coefficients of the order $({\alpha}/{4\pi}) \, C^{e\mu ee}_{\ell e} \log(m_Z/m_\mu)$. This explains why the flat direction of the orange region corresponds to $C^{e\mu ee}_{\ell e}(\Lambda) \approx 10^2 \times C^{(1)\,e\mu}_{\vp \ell}(\Lambda)$. 
We can also see that $\mu\to eee$ is only mildly suppressed (along another direction), which implies a combined indirect constraint of about $\text{BR}(Z\to \mu e)\lesssim 10^{-14}$. More in general, no exact cancellation can occur in presence of only two dominant WCs contributing to $\mu\to eee$. The reason is that, as one can see from the formulae in \aref{App:3body}, a strong suppression of $\mu\to eee$ would require a simultaneous cancellation of the coefficients in Eqs.~(\ref{eq:CVLR},\,\ref{eq:CVRL}), which is only possible if the coefficient of the RH Higgs-lepton operator $C^{(1)\,e\mu}_{\vp e}$ is also tuned to cancel out
$C^{e\mu ee}_{\ell e}$ in Eq.~(\ref{eq:CVRL}). Needless to say, yet another tuning would be required to suppress $\mu\to e$ conversion, such as in the top panel of the figure.\footnote{Moreover, the above-discussed operators involved in $\mu \to eee$ would destabilise the cancellation of $\mu\to e$ conversion and a fine adjustment of the flat direction shown in the top panel would be needed.}

In summary, a simultaneous cancellation of $\mu\to eee$ and $\mu\to e$ conversion in nuclei able to enhance the maximum possible BR$(Z\to \mu e)$ would require a fine tuning involving at least four operators and thus it looks extremely unlikely in the context of any UV-complete model. Therefore, we can conclude that we do not expect $Z\to \mu e$ to be observed at the CEPC or FCC-ee even if the optimal sensitivity shown in Table~\ref{Table:Zdecay_limits} ($10^{-10}$) could be reached.
Incidentally, note the astonishing sensitivity of the upcoming low-energy LFV experiments Mu3e, Mu2e and COMET, which will constrain the coefficients of certain operators down to even $10^{-7}$-$10^{-8}$ if LFV effects are induced by TeV-scale new physics.


\begin{figure}[t!]
\centering
\includegraphics[width=0.9\textwidth]{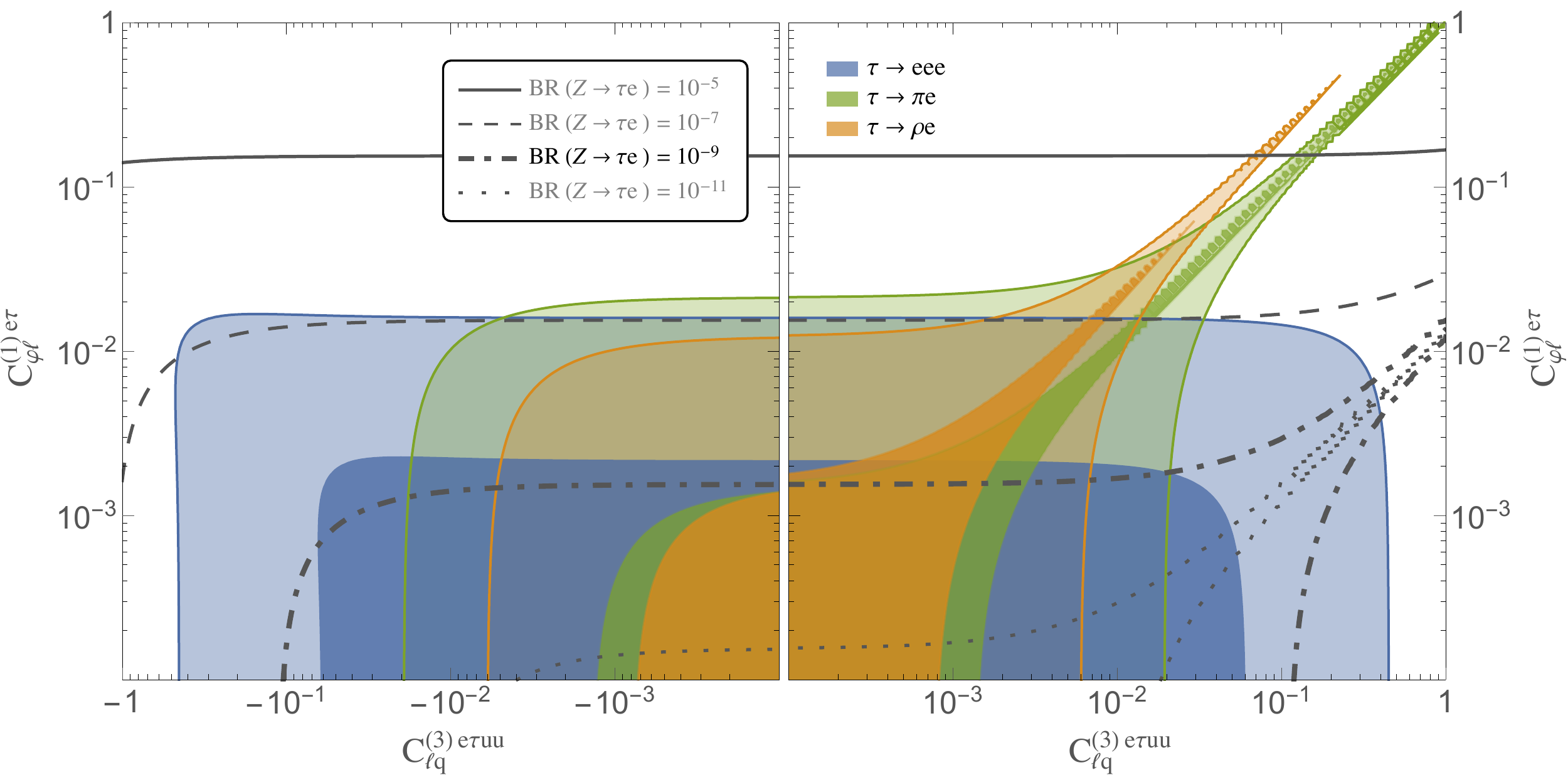}
\includegraphics[width=0.9\textwidth]{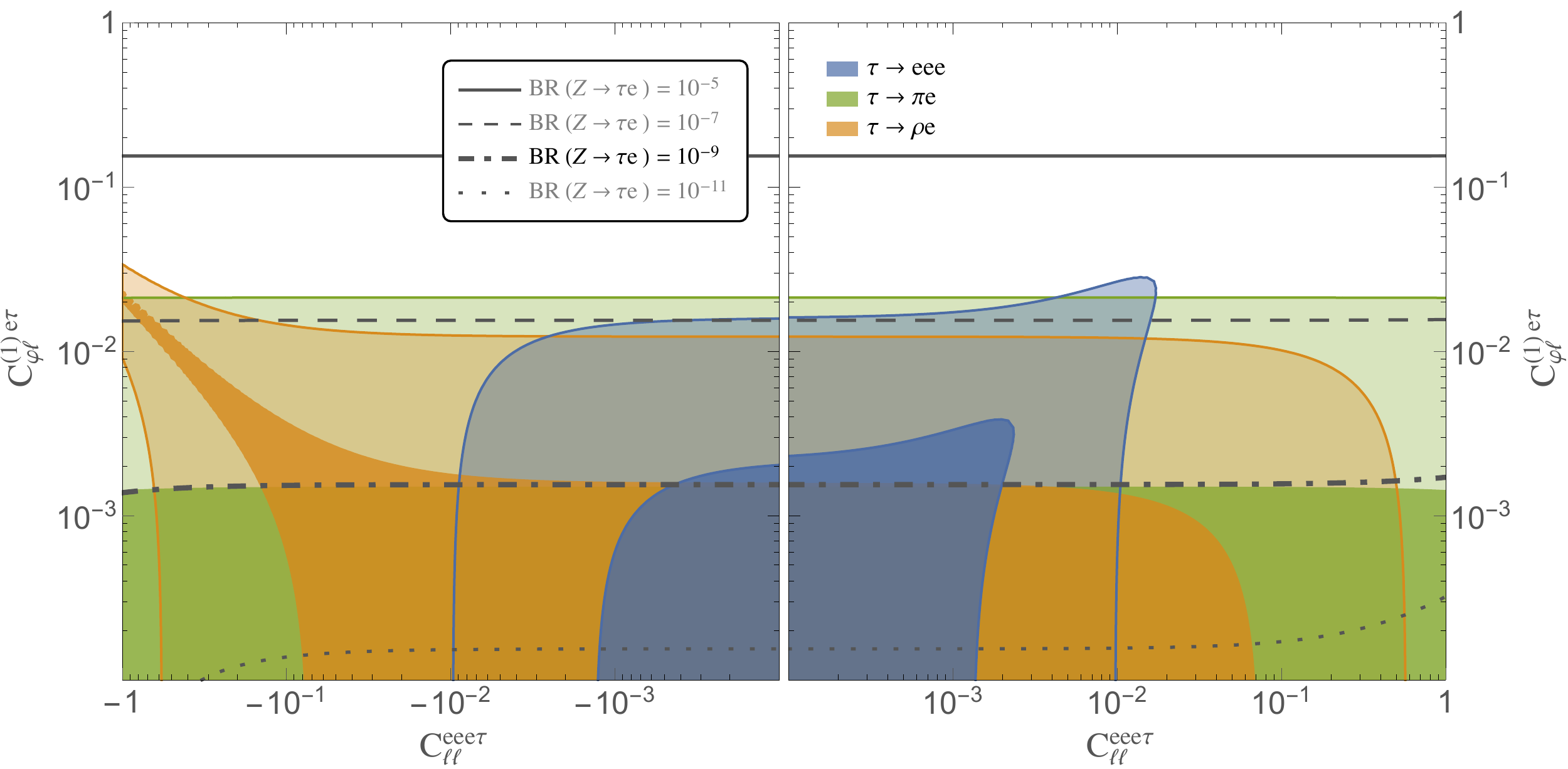}
\caption{
Contours of BR($Z\to \tau e$) as a function of 
$C^{(1)\,e\tau}_{\vp \ell}(\Lambda)$ and $C^{(3)\,e\tau uu}_{\ell q}(\Lambda)$ (top panel), $C^{(1)\,e\tau}_{\vp \ell}(\Lambda)$ and $C^{ee e\tau}_{\ell \ell}(\Lambda)$ (bottom panel), highlighting the future Tera Z sensitivity of $10^{-9}$. The cutoff scale is set to $\Lambda = 1$~TeV. The lighter coloured regions are currently allowed by $\tau\to eee$ (blue), $\tau\to \pi e$ (green), $\tau\to \rho e$ (orange), the corresponding darker regions show the future expected sensitivities.
\label{fig:results-taue-1}}
\end{figure}

\begin{figure}[t!]
\centering
\includegraphics[width=0.9\textwidth]{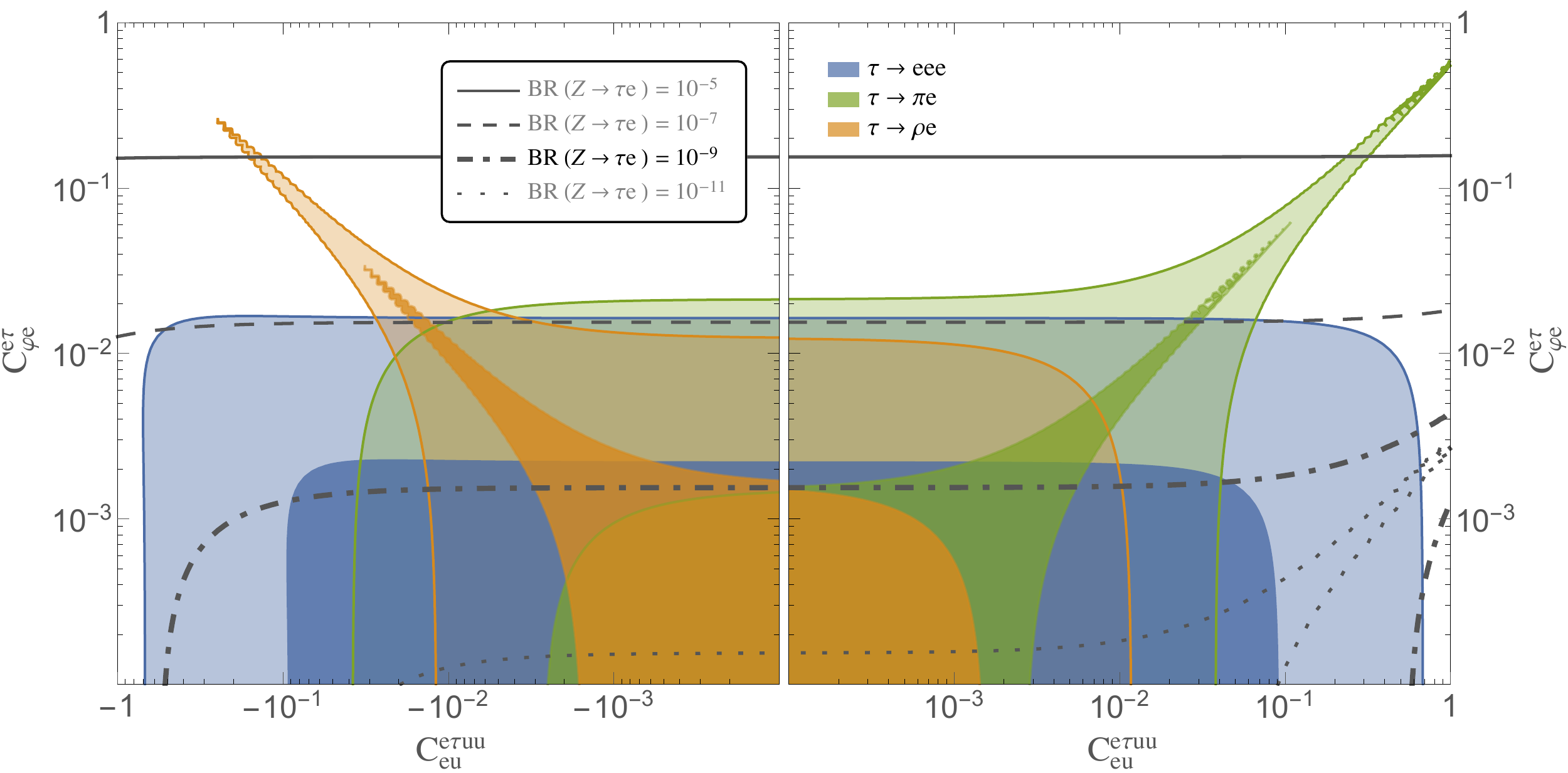}
\caption{
Same as \fref{fig:results-taue-1} for  $C^{e\tau}_{\vp e}(\Lambda)$ vs $C^{e\tau uu}_{e u}(\Lambda)$.
\label{fig:results-taue-2}}
\end{figure}

\paragraph{$\boldsymbol{\tau-\ell}$ sector.}
In Figures~\ref{fig:results-taue-1} and~\ref{fig:results-taue-2}, we show our results for $Z\to  \tau e$ and the LFV $\tau$ decays $\tau\to eee$, $\tau \to \pi e$, and $\tau \to \rho e$, bearing in mind that the corresponding plots for the $\tau-\mu$ sector are virtually the same once the obvious changes ({\it e.g.}~of flavour indices) are taken into account.
\fref{fig:results-taue-1} displays the possible interference between the $C^{(1)\,e\tau}_{\vp \ell}$ and the coefficient of the 2-lepton 2-quark operator $C^{(3)\,e\tau uu}_{\ell q}$~(top), and the 4-lepton operator $C^{ee e\tau}_{\ell \ell}$~(bottom).
As in the case of the $\mu-e$ sector, we observe that the WCs could conspire to cancel out 
BR$(Z\to  \tau e)$, a possibility that is however incompatible with the low-energy constraints. 

Conversely, \fref{fig:results-taue-1} shows that $\tau \to \pi e$ and $\tau \to \rho e$ can be suppressed, but in general not simultaneously. This is due to the different dependence of the amplitudes of the two semileptonic decays on the WCs. For instance, from the top panel, we see that  $\tau \to \pi e$  vanishes for $C^{(3)\,e\tau uu}_{\ell q} (\Lambda) \approx C^{(1)\,e\tau}_{\vp \ell}(\Lambda)$, while a cancellation of  $\tau \to \rho e$ requires $C^{(3)\,e\tau uu}_{\ell q} (\Lambda) \approx 0.5 \times C^{(1)\,e\tau}_{\vp \ell}(\Lambda)$. Both numerical values are well accounted for by the approximate formulae in Appendices~\ref{App:semileptonicV} and~\ref{App:semileptonicP}, see Eqs.~(\ref{eq:rho-flatD},\,\ref{eq:pi-flatD}).
This result implies that it is not possible to tune the parameters to cancel both semileptonic $\tau$ decays and enhance the LFVZD effects. This is also shown clearly in the bottom panel of the figure, where $\tau \to \rho e$ features a flat direction while $\tau \to \pi e$ does not. 
Notice that the latter is actually almost independent of $C^{ee e\tau}_{\ell \ell}$, since the RGE contributions of the vectorial 4-lepton operators are very suppressed for this channel involving a pseudoscalar meson.

For both cases depicted in~\fref{fig:results-taue-1}, we can see that suppressing the purely leptonic LFV decays such as $\tau \to eee$ would require introducing and tuning additional operators, hence it is difficult to envisage a situation where BR$(Z\to \tau e)$ can attain values much larger than the limits we obtained in~\tref{Table:Single operators bound} for the single operator analysis. Still, note that, once the combined constraints are fulfilled, BR$(Z\to  \tau e)$ can be as large as $\approx 10^{-7}$, a rate which is largely within the sensitivity of a Tera Z factory, $\approx 10^{-9}$, see~\tref{Table:Zdecay_limits}. 
Consequently, the plots show the nice complementarity in testing the NP parameter space of LFVZD searches at a Tera Z and the prospected sensitivity on LFV $\tau$ decays at Belle-II.

So far we have considered the LH Higgs-lepton operator $C^{(1)\,e\tau}_{\vp \ell}$ (the results for $C^{(3)\,e\tau}_{\vp \ell}$ being almost identical). \fref{fig:results-taue-2} shows that the outcome of our analysis does not qualitatively change if we consider operators involving RH currents: in this example, $C^{e\tau}_{\vp e}$ and the 2-lepton 2-quark operator $C^{e\tau uu}_{e u}$. The different behaviours of $\tau \to \pi e$ and $\tau \to \rho e$ are even more striking due to a relative sign difference of the $C^{e\tau}_{\vp e}$ contributions, see Eqs.~(\ref{eq:rho-flatD-RH},\,\ref{eq:pi-flatD-RH}), accentuating the complementarity between different LFV observables.
As before, we find maximum allowed LFVZD rates well within the expected sensitivities at the Tera Z factory.

To summarise, also for the observables involving $\tau$ leptons, we found that accidental cancellations of either $Z\to \tau\ell$ or $\tau$ LFV decays are extremely unlikely. Moreover, Figures~\ref{fig:results-taue-1} and~\ref{fig:results-taue-2} show that, if the UV physics induces several operators simultaneously with coefficients of comparable size (let's say, $10^{-3}/(1~\text{TeV})^2$) including one or more lepton-Higgs operators, the capability of testing new physics of a Tera Z factory through searches for $Z\to \tau\ell$ is comparable to or better than the one of Belle-II through $\tau$ LFV processes. 
\section{Summary and Conclusions}
\label{Sec:concl}

In this work we assessed the potential of the proposed circular $e^+e^-$ colliders CEPC and FCC-ee to discover new physics by searching for the LFV decays $Z\to \ell_i \ell_j$ with data collected operating at a center-of-mass energy $\approx m_Z$, and under the assumption that an integrated luminosity corresponding to
$\mathcal{O}(10^{12})$  visible $Z$ decays (``Tera Z'') will be reached.
As shown in \tref{Table:Zdecay_limits}, a Tera Z factory has the potential to improve the present limits by more than three orders of magnitude. 
It is however not obvious whether such an excellent sensitivity is enough to test or discover new physics effects, since any new dynamics responsible for LFV interactions of the $Z$ bosons would unavoidably induce LFV decays of muons and taus, which in turn set indirect limits on the maximum possible rates of $Z\to \ell_i \ell_j$. 
We evaluated these indirect constraints within the model-independent framework of the SMEFT.

The main findings of our analysis can be summarised as follows.
\begin{itemize}
\item Of the five SMEFT operators that can induce LFVZD at tree level shown in~\tref{Tab:dim6}, only three, belonging to the class of the lepton-Higgs operators can lead to observable effects. The other two, that is, the dipole operators, are too tightly constrained by the low-energy LFV radiative decays $\ell_i\to \ell_j\gamma$, see \tref{Table:Single operators bound} and \fref{Fig:Single operators dominance}.
\item Combining the current bounds on BR$(\mu\to eee)$ and on the $\mu\to e$ conversion rate in atomic nuclei results in a strong indirect constraint on BR$(Z\to \mu e)$. The distinct dependence of the two processes  (and also of the conversion rates in different nuclei) on LFV 4-fermion operators makes accidental cancellations that could enhance the allowed $Z\to \mu e$ rate very unlikely. This would indeed require a careful tuning of the coefficients of at least four independent operators. As a consequence, in a realistic UV model, the current (future) low-energy bounds imply BR$(Z\to \mu e)\lesssim 10^{-12}~(10^{-16})$ (see \fref{fig:results-mue}), way beyond the expected sensitivity of Tera Z factories.   
\item Also in the case of $\tau-\ell$ transitions, the interplay of distinct LFV $\tau$ decays (in particular $\tau\to \rho \ell$ and $\tau \to \pi \ell$ that, despite involving the same fermions, have a prominently different dependence on LFV operators) exclude the possibility that low-energy constraints can be realistically tuned away, cf.~Figs.~\ref{fig:results-taue-1} and~\ref{fig:results-taue-2}.
\item Despite the constraints from low-energy LFV processes, BR($Z\to\tau\ell$) is still allowed to be as large as $\approx 10^{-7}$, more than one order of magnitude below the current LHC bounds (and thus  beyond the reach of HL-LHC too), but well within the sensitivity of future Tera Z factories,~$\approx 10^{-9}$, cf.~\tref{Table:Zdecay_limits}. This is clearly a  consequence of the fact that $\tau$ LFV processes are comparatively less constrained than the muon ones. 
\item Even considering the most optimistic future bounds from Belle-II, we showed in Figs.~\ref{fig:results-taue-1} and~\ref{fig:results-taue-2} that Tera-Z searches for LFV $Z$ decays will be able to test new physics as effectively as LFV $\tau$ decays (and up to scales of the order of 20-30 TeV as shown in~\fref{Fig:Single operators dominance}). 
This is the case even if the UV new physics gives rise not only to the lepton-Higgs operators responsible for $Z\to \tau \ell$, but also directly to 4-fermion operators, as long as the coefficients of the latter are of the same order as those of the former (or smaller). On the other hand, in order for the LFV $Z$ decays to be observable, 
the coefficients of the dipole operators need to be somewhat suppressed (which is certainly possible, since they are unavoidably generated at loop level).
\end{itemize}

In summary, we have shown that a future Tera Z factory has the potential to probe new LFV physics in the $\tau-\ell$ sector at the same level of other low-energy experiments. 
They can thus not only provide new independent insight into these phenomena, but also play a major role if Belle-II does not reach the most optimistic background-free environment.
Finally, it is worth stressing the importance of the $Z\to \mu e$ searches even if they are disfavoured by our analysis, as their experimental observation would imply a new physics scenario even beyond the SMEFT framework.  

\medskip
\paragraph{Acknowledgments.}
We would like to thank Claudia Garc\'ia-Garc\'ia, Tong Li and Michael Schmidt for very valuable discussions.
This project received support from the DFG Collaborative Research Center SFB1258. 
L.C.~is partially supported by the National Natural Science Foundation of China under the grant No.~12035008. X.M.~is supported by the Alexander von Humboldt Foundation.


\appendix

\section{Tree-level expressions for low-energy LFV observables}\label{App:lowLFV}

We collect here the analytical expressions for the low-energy LFV observables we considered in our analysis. 
We are interested in expressing them directly in terms of the SMEFT parameters, therefore when the available results in the literature are given only in the LEFT basis, we translate them applying the tree-level matching of Ref.~\cite{Jenkins:2017jig}.
In the case of the 3-body decays $\ell_i\to\ell_j\ell_k\bar\ell_m$, however, we computed and compared them to previous results. 

Notice that these expressions neglect the effect of both high and low-energy RGE, which we do include in our numerical analysis by means of the {\it wilson}~\cite{Aebischer:2018bkb} and {\it flavio}~\cite{Straub:2018kue} packages. 
Nevertheless, we find that the following expressions are helpful to understand analytically the overall behaviour of the observables. 

\subsection[{Radiative decays $\ell_i \to \ell_j\gamma$}]{Radiative decays $\boldsymbol{\ell_i \to \ell_j\gamma}$}

The branching ratio of radiative lepton decay  processes can be expressed in terms of two form factors $F_{TL}$ and $F_{TR}$~\cite{Crivellin:2013hpa}:
\begin{equation}
{\rm BR} \left(\ell_{i} \to \ell_{j}
  \gamma\right)\,=\,\dfrac{m_{\ell_i}^3}{16\pi\Lambda^4 \,
  \Gamma_{\ell_i}} \left( \big|F^{ji}_{TR} \big|^2 + \big|F^{ji}_{TL} \big|^2\right ) \,.
\label{Brmuegamma}
\end{equation}
At tree-level, these form factors match to the SMEFT photon dipole defined in Eq.~\eqref{SMEFTphotondipole} as
\bea
F_{TR}^{ji} = F_{TL}^{ij\,*} =  v\sqrt{2} C_{e\gamma}^{ji}\,.
\label{eq:phtree}
\eea
Additional contributions beyond tree-level have also been computed, which could be important when the photon dipole is not generated at tree level. 
See for instance Refs.~\cite{Crivellin:2013hpa,Pruna:2014asa,Crivellin:2017rmk} for further details.

\subsection[Leptonic 3 body decays  $\ell_i \to \ell_j\ell_k\bar\ell_m$]{Leptonic 3 body decays  $\boldsymbol{\ell_i \to \ell_j\ell_k\bar\ell_m}$}
\label{App:3body}

Given some small discrepancies between different available computations~\cite{Kuno:1999jp,Brignole:2004ah,Crivellin:2013hpa,Crivellin:2017rmk}, we compute again the decay rates for $\ell_i \to \ell_j\ell_k\bar\ell_m$.
Starting from the LEFT Lagrangian~\cite{Jenkins:2017jig}, the relevant terms for the tree-level 3 body decays are
\begin{align}
\mathcal L_{\rm LEFT} &\supset
C_{ee}^{VLL}\, \big(\bar e_j \gamma^\mu P_L e_i\big) \big(\bar e_k \gamma_\mu P_L e_m\big)
+C_{ee}^{VRR}\, \big(\bar e_j \gamma^\mu P_R e_i\big) \big(\bar e_k \gamma_\mu P_R e_m\big) \nonumber\\
&+ C_{ee}^{VLR}\, \big(\bar e_j \gamma^\mu P_L e_i\big) \big(\bar e_k \gamma_\mu P_R e_m\big)
+ \Big\{ C_{ee}^{SRR} \big(\bar e_j P_R e_i\big)\big(\bar e_k P_R e_m\big) + h.c. \Big\} \nonumber\\
&+ \Big\{ C_\gamma \big(\bar e_j \sigma^{\mu\nu} P_R e_i\big) F_{\mu\nu} + h.c. \Big\}\,.
\end{align} 
The expressions for the decays depend on the flavor combinations of the final leptons, as they could involve new possible contractions and symmetry factors. 
For the $\mu\to eee$, $\tau\to eee$ and $\tau\to\mu\mu\mu$ decays we get,
\begin{align}
{\rm BR}(\ell_i\to \ell_j \ell_j \bar \ell_j) &
=\frac{m_{\ell_i}^5}{3 (16\pi)^3 \Gamma_{\ell_i}} 
\bigg\{16\big|C^{VLL}\big|^2+ 16 \big|C^{VRR}\big|^2 + 8 \big|C^{VLR}\big|^2 + 8\big|C^{VRL}\big|^2 \nn&
+ \big|C^{SRR}\big|^2 + \big|C^{SLL}\big|^2 
+ \frac{256 e^2}{m_{\ell_i}^2} \left(\log\frac{m_{\ell_i}^2}{m_{\ell_j}^2}-\frac{11}4\right)
\Big(\big|C_{\gamma}^{ji}\big|^2 + \big|C_\gamma^{ij}\big|^2\Big)\nn
&- \frac{64 e}{m_{\ell_i}}\, Re\Big[\Big(2 C^{VLL} + C^{VLR}\Big) C_\gamma^{ji\, *} + \Big(2C^{VRR}+C^{VRL}\Big) C_\gamma^{ij}\Big] \bigg\}\,,
\end{align}

with the matching conditions given by
\begin{align}
C^{VLL} &= \frac1{\Lambda^2}\Big\{(2\sw^2-1) \left( C_{\vp \ell}^{(1)ji} + C_{\vp\ell}^{(3)ji} \right) + C_{\ell\ell}^{jijj}\Big\}\,, \\
C^{VRR} &= \frac1{\Lambda^2}\Big\{2 \sw^2 C_{\vp e}^{ji} +C_{ee}^{jijj} \Big\}\,,\\ 
C^{VLR} &=  \frac1{\Lambda^2}\Big\{2\sw^2 \left( C_{\vp\ell}^{(1)ji} + C_{\vp \ell}^{(3)ji} \right) + C_{\ell e}^{jijj}\Big\}\,,\label{eq:CVLR}\\
C^{VRL} &=  \frac1{\Lambda^2}\Big\{(2\sw^2-1) C_{\vp e}^{ji} + C_{\ell e}^{jjji}\Big\}\,, \label{eq:CVRL}\\
C^{SRR} &= C^{SLL} = 0\,,
\end{align}
and the photon dipole matching given in Eq.~\eqref{SMEFTphotondipole}.
Similarly, for the decays $\tau\to e\mu\bar\mu$ and $\tau\to \mu e\bar e$, we find, 
\begin{align}
{\rm BR}(\ell_i\to \ell_j \ell_k \bar \ell_k) &
=\frac{m_{\ell_i}^5}{3 (16\pi)^3 \Gamma_{\ell_i}} 
\bigg\{8\big|C^{VLL}\big|^2+ 8 \big|C^{VRR}\big|^2 + 8 \big|C^{VLR}\big|^2 + 8\big|C^{VRL}\big|^2 \nn&
+ 2\big|C^{SRR}\big|^2 + 2\big|C^{SLL}\big|^2 
+ \frac{256 e^2}{m_{\ell_i}^2} \left(\log\frac{m_{\ell_i}^2}{m_{\ell_k}^2}-3\right)
\Big(\big|C_{\gamma}^{ji}\big|^2 + \big|C_\gamma^{ij}\big|^2\Big)\nn
&- \frac{64 e}{m_{\ell_i}}\, Re\Big[\Big(C^{VLL} + C^{VLR}\Big) C_\gamma^{ji\, *} + \Big(C^{VRR}+C^{VRL}\Big) C_\gamma^{ij}\Big] \bigg\}\,,
\end{align}
with
\begin{align}
C^{VLL} &= \frac1{\Lambda^2}\Big\{(2\sw^2-1) \left( C_{\vp \ell}^{(1)ji} + C_{\vp\ell}^{(3)ji} \right) + C_{\ell\ell}^{jikk}\Big\}\,, \\
C^{VRR} &= \frac1{\Lambda^2}\Big\{2 \sw^2 C_{\vp e}^{ji} +C_{ee}^{jikk}\Big\} \,,\\ 
C^{VLR} &=  \frac1{\Lambda^2}\Big\{2\sw^2 \left( C_{\vp\ell}^{(1)ji} + C_{\vp \ell}^{(3)ji} \right) + C_{\ell e}^{jikk}\Big\}\,,\\
C^{VRL} &=  \frac1{\Lambda^2}\Big\{(2\sw^2-1) C_{\vp e}^{ji} + C_{\ell e}^{kkji}\Big\}\,, \\
C^{SRR} &= C^{SLL} = 0\,,
\end{align}
Finally, for $\tau\to e e \bar\mu$ and $\tau\to \mu\mu\bar e$, 
\begin{align}
{\rm BR}(\ell_i\to \ell_k \ell_k \bar \ell_j) &
=\frac{m_{\ell_i}^5}{3 (16\pi)^3 \Gamma_{\ell_i}} 
\bigg\{16\big|C^{VLL}\big|^2+ 16 \big|C^{VRR}\big|^2 + 8 \big|C^{VLR}\big|^2 + 8\big|C^{VRL}\big|^2 \nn&
\hspace{3cm}+ \big|C^{SRR}\big|^2 + \big|C^{SLL}\big|^2 \bigg\}
\end{align}
with 
\begin{align}
C^{VLL} &= C_{\ell\ell}^{kikj}/\Lambda^2\,, \\
C^{VRR} &=C_{ee}^{kikj}/\Lambda^2 \,,\\ 
C^{VLR} &= C_{\ell e}^{kikj}/\Lambda^2\,,\\
C^{VRL} &=  C_{\ell e}^{kjki}/\Lambda^2\,, \\
C^{SRR} &= C^{SLL} = 0\,.
\end{align}
Our results are in agreement, when comparison is possible, to those of~\cite{Kuno:1999jp,Brignole:2004ah,Crivellin:2017rmk}.
\subsection[Semileptonic $\tau$ decays  $\tau \to \mesonV \ell$]{Semileptonic $\boldsymbol\tau$ decays  $\boldsymbol{\tau \to \mesonV \ell}$}
\label{App:semileptonicV}

The expressions for $\tau\to \mesonV\ell$, with $\ell=e,\mu$ and $\mesonV=\rho, \phi$ a vectorial meson, in terms of the LEFT Wilson coefficients can be found in Ref.~\cite{Aebischer:2018iyb}. 
In order to express them in terms of the SMEFT parameters, we use the tree-level matching relations from Ref.~\cite{Jenkins:2017jig}.
Here we report the resulting expressions. 

The branching ratio can be expressed as
\be\label{GammaSemilep}
{\rm BR} (\tau\to \mesonV \ell) = \frac{\sqrt{\lambda(m_\tau^2,m_\ell^2,m_\mesonV^2)}}{16\pi m_\tau^3\,\Gamma_\tau}\, \overline{\big|\mathcal M^{\phantom V}_{\tau\to \mesonV\ell}\big|}^{\,2}\,,
\ee
with $\lambda(a,b,c)=a^2+b^2+c^2-2(ab+ac+bc)$, and the amplitude $\mathcal M$ receiving contributions of both vectorial and tensorial leptonic currents:
\be
\mathcal M_{\tau\to\mesonV\ell} = \mathcal M_{\tau\to\mesonV\ell}^{\rm VC} +\mathcal M_{\tau\to\mesonV\ell}^{\rm TC}\,.
\ee
From here, the squared amplitude is given as
\be
 \overline{\big|\mathcal M^{\phantom V}_{\tau\to \mesonV\ell}\big|}^{\,2} =  
 \overline{\big|\mathcal M^{\rm VC}_{\tau\to \mesonV\ell}\big|}^{\,2} +  \overline{\big|\mathcal M^{\rm TC}_{\tau\to \mesonV\ell}\big|}^{\,2}+ \mathcal I_{\,\tau\to \mesonV\ell}\,,
 \ee
 with
\begin{align}
 \overline{\big|\mathcal M^{\rm VC}_{\tau\to \mesonV\ell}\big|}^{\,2} &=
 \frac12\bigg\{\Big(\big|\gVLV\big|^2+\big|\gVRV\big|^2\Big) \bigg(\frac{(m_\tau^2-m_\ell^2)^2}{m_\mesonV^2} + m_\tau^2+m_\ell^2-2m_\mesonV^2\bigg) \nonumber \\
 &\hspace{1.cm}  -12 m_\tau m_\ell\, \Re\Big[\gVRV \big(\gVLV\big)^*\Big]
 \bigg\}\,, \\
  \overline{\big|\mathcal M^{\rm TC}_{\tau\to \mesonV\ell}\big|}^{\,2} &=
 \frac12\bigg\{\Big(\big|\gTLV\big|^2+\big|\gTRV\big|^2\Big) \Big(2\big(m_\tau^2-m_\ell^2\big)^2 - m_\mesonV^2(m_\tau^2+m_\ell^2) - m_\mesonV^4\Big)\nonumber \\
 &\hspace{1.cm}  -12 m_\mesonV^2m_\tau m_\ell\, \Re\Big[\gTRV \big(\gTLV\big)^*\Big]
 \bigg\}\,, 
 \end{align}
and the interference
\begin{align}
\mathcal I_{\,\tau\to \mesonV\ell} &= 
3 m_\tau (m_\tau^2-m_\ell^2-m_{\mesonV}^2)\, \Re\Big[\gVLV \big(\gTRV\big)^* + \gVRV\big(\gTLV\big)^*\Big]\nonumber\\
&+ \hspace{.05cm}3 m_\ell\hspace{.05cm}(m_\ell^2-m_\tau^2-m_{\mesonV}^2)\, \Re\Big[\gVRV \big(\gTRV\big)^* + \gVLV\big(\gTLV\big)^*\Big]\,.
\end{align}
After matching, the vectorial current receives contributions from the 2 lepton-2 quark vectorial operators and from the Higgs-lepton ones, while the tensorial current gets contributions from the dipoles and the tensorial $Q_{\ell e qu}^{(3)}$ in \tref{Tab:dim6}.
The final expressions depend on the meson and for $\mesonV=\rho, \phi$ they are given by
\begin{align}
\label{eq:rho-flatD}
\gVLV[\rho] &= \frac{m_\rho f_\rho}{\sqrt2 \Lambda^2}  \left\{ (1-2\sw^2)\Big(C_{\vp l}^{(1)} + C_{\vp l}^{(3)}\Big)^{\tau\ell} - C_{\ell q}^{(3)\, \tau\ell u u} +\frac12 C_{\ell u}^{\tau\ell uu}-\frac12 C_{\ell d}^{\tau\ell d d}\right\}\,,\\
\label{eq:rho-flatD-RH}
\gVRV[\rho]&= \frac{m_\rho f_\rho}{\sqrt2 \Lambda^2}  \left\{ (1-2\sw^2)\, C_{\vp e}^{\tau\ell} +\frac12 C_{e u}^{\tau\ell uu}-\frac12 C_{e d}^{\tau\ell d d}\right\}\,,\\
\gVLV[\phi] &= \frac{m_\phi f_\phi}{2 \Lambda^2}  \left\{ \Big(\frac43\sw^2-1\Big)\Big(C_{\vp l}^{(1)} + C_{\vp l}^{(3)}\Big)^{\tau\ell} + C_{\ell q}^{(1)\, \tau\ell ss} + C_{\ell q}^{(3)\, \tau\ell ss} + C_{\ell d}^{\tau\ell ss}\right\}\,,\\
\gVRV[\phi]&= \frac{m_\phi f_\phi}{2 \Lambda^2}  \left\{ \Big(\frac43\sw^2-1\Big)\, C_{\vp e}^{\tau\ell} + C_{e d}^{\tau\ell ss}+ C_{qe}^{ss\tau\ell}\right\}\,,\\
\gTLV[\rho] &= -\frac1{\Lambda^2} \left\{\frac{ev f_\rho}{m_\rho}\, C_\gamma^{\tau\ell} + \sqrt2 f_{T\rho}\, C_{\ell equ}^{(3)\, \tau\ell uu}\right\}^*\,,\\
\gTRV[\rho] &= -\frac1{\Lambda^2} \left\{\frac{ev f_\rho}{m_\rho}\, C_\gamma^{\ell\tau} + \sqrt2 f_{T\rho}\, C_{\ell equ}^{(3)\, \ell\tau uu}\right\}\,,\\
\gTLV[\phi] &= \frac{\sqrt2 ev f_\phi}{3m_\phi\Lambda^2}\, C_{e\gamma}^{\tau\ell\, *} \,,\\
\gTRV[\phi] &= \frac{\sqrt2 ev f_\phi}{3m_\phi\Lambda^2}\, C_{e\gamma}^{\ell\tau} \,,
\end{align}
with $f_{(T)\mesonV}$ the (transverse) decay constants of the vectorial mesons.

\subsection[Semileptonic $\tau$ decays  $\tau \to \mesonP \ell$]{Semileptonic $\boldsymbol\tau$ decays  $\boldsymbol{\tau \to \mesonP \ell}$}
\label{App:semileptonicP}

Equivalently to the vectorial mesons, we take the expressions for the decays to a pseudoscalar meson $\mesonP$ from Ref.~\cite{Aebischer:2018iyb} and apply the tree-level matching relations from Ref.~\cite{Jenkins:2017jig}.

The branching ratio for $\tau\to \mesonP\ell$, with $\ell=e,\mu$ and $\mesonP=\pi^0, K^0$, is given again by Eq.~\eqref{GammaSemilep} after replacing $\mesonV$ by $\mesonP$. 
The squared decay amplitude in this case can be expressed as, 
\be
 \overline{\big|\mathcal M_{\tau\to \mesonP\ell}\big|}^{\,2} =  
 \frac12\big(m_\tau^2+m_\ell^2-m_{\mesonP}^2\big)\,\Big(\big|\gLP\big|^2 + \big|\gRP\big|^2\Big)
 +2m_\tau m_\ell \, \Re\Big[\gLP \big(\gRP\big)^*\Big]\,,
\ee
with 
\be
\gLP = \gSLP - m_\ell \gVLP + m_\tau \gVRP\,,
\qquad
\gRP = \gSRP - m_\ell \gVRP + m_\tau \gVLP\,.
\ee
The expressions for the scalar and vectorial couplings depend on the meson and for $\mesonP=\pi^0, K^0$ they are given by
\begin{align}
\label{eq:pi-flatD} 
\gVLP[\pi] &= \frac{f_\pi}{2\sqrt2 \Lambda^2} \left\{ -2 \Big(C_{\vp l}^{(1)} + C_{\vp l}^{(3)}\Big)^{\tau\ell}
+ C_{\ell u}^{\tau\ell uu} - C_{\ell d}^{\tau\ell dd} + 2 C_{\ell q}^{(3)\, \tau\ell uu} \right\}\,, \\
\label{eq:pi-flatD-RH}
\gVRP[\pi] &= \frac{f_\pi}{2\sqrt2 \Lambda^2} \left\{ -2 C_{\vp e}^{\tau\ell} + C_{eu}^{\tau\ell uu}- C_{ed}^{\tau\ell dd}\right\}\,,\\
\gVLP[K] & = \frac{f_K}{2\Lambda^2}\, \bigg\{ C_{\ell q}^{(1)\, \ell\tau ds} +C_{\ell q}^{(3)\, \ell\tau ds} - C_{\ell d}^{\ell\tau ds}\bigg\}\,,\\
\gVRP[K] & = \frac{f_K}{2\Lambda^2}\, \bigg\{ C_{qe}^{ds\ell\tau} - C_{ed}^{\ell\tau ds}\bigg\}\,,\\
\gSLP[\pi] &= \frac{f_\pi m_\pi^2}{2\sqrt2 (m_u+m_d) \Lambda^2}\, \left\{ C_{\ell equ}^{(1)\, \tau\ell uu} - C_{\ell edq}^{\tau\ell dd} \right\}^*\,,\\
\gSRP[\pi] &= \frac{f_\pi m_\pi^2}{2\sqrt2 (m_u+m_d) \Lambda^2}\, \left\{ C_{\ell edq}^{\ell\tau dd} - C_{\ell equ}^{(1)\, \ell\tau uu} \right\}\,,\\
\gSLP[K] & = - \frac{f_K m_K^2}{2(m_s+m_d) \Lambda^2}\, C_{\ell edq}^{\tau\ell sd\, *}\,,\\
\gSRP[K] & =  \frac{f_K m_K^2}{2(m_s+m_d) \Lambda^2}\, C_{\ell edq}^{\ell\tau ds}\,,
\end{align}
with $f_{\mesonP}$ the decay constants of the pseudoscalar mesons.

\subsection[$\mu\to e$ conversion in nuclei]{$\boldsymbol{\mu\to e}$ conversion in nuclei}
\label{App:mueconversion}

The conversion rate can be expressed as~\cite{Kitano:2002mt},
\be
{\rm CR} (\mu-e, N) = \frac1{\Gamma_{\rm capt}}\frac{m_\mu^5}{\Lambda^4}\bigg\{
\Big| \widetilde C_{DL} D + \widetilde C_{SL}^{(p)} S^{(p)} + \widetilde C_{SL}^{(n)} S^{(n)} + \widetilde C_{VL}^{(p)} V^{(p)} + \widetilde C_{VL}^{(n)} V^{(n)} \Big|^2
+\Big| L\leftrightarrow R \Big|^2 \bigg\}\,,
\label{eq:mueconv}
\ee
with $\Gamma_{\rm capt}$ being the muon capture rate in a nuclei $N$ and
$D, S^{(p/n)}$ and $V^{(p/n)}$ the dimensionless overlap integrals, whose numerical values depend on the nuclei and can be found in Ref.~\cite{Kitano:2002mt}.
After tree-level matching~\cite{Jenkins:2017jig}, we obtain that the dipole form factors are given by
\be
\widetilde C_{DL} = \frac v{2\sqrt2 m_\mu}\, C_{e\gamma}^{e\mu}\,, \qquad  
\widetilde C_{DR} = \frac v{2\sqrt2 m_\mu}\, C_{e\gamma}^{\mu e\, *}\,,
\ee
the vector form factors by
\be
\widetilde C_{VL}^{(p)} = 2 C_{VX}^{(u)} + C_{VX}^{(d)}\,,\qquad
\widetilde C_{VL}^{(n)} = C_{VX}^{(u)} + 2 C_{VX}^{(d)}\,,
\ee
with
\begin{align}
C_{VL}^{(u)} &=  \Big(C_{\ell q}^{(1)} - C_{\ell q}^{(3)} + C_{\ell u}\Big)^{e\mu uu} + \Big(1-\frac83 \sw^2\Big) \Big(C_{\vp \ell}^{(1)} + C_{\vp \ell}^{(3)}\Big)^{e\mu}\,,\\
C_{VL}^{(d)} &= \Big(C_{\ell q}^{(1)} + C_{\ell q}^{(3)} + C_{\ell d}\Big)^{e\mu dd} -\Big(1-\frac43 \sw^2\Big) \Big(C_{\vp \ell}^{(1)} + C_{\vp \ell}^{(3)}\Big)^{e\mu}\,,\\
C_{VR}^{(u)} &= C_{e u}^{e\mu uu} +C_{qe}^{uue\mu} + \Big(1-\frac83 \sw^2\Big) C_{\vp e}^{e\mu}\,,\\
C_{VR}^{(d)} &= C_{e d}^{e\mu dd} +C_{qe}^{dde\mu} - \Big(1-\frac43 \sw^2\Big) C_{\vp e}^{e\mu}\,,
\end{align}
and finally the scalar form factors by
\begin{align}
\widetilde C_{SL}^{(p/n)} &= - G_S^{(u,p/n)}\, C_{\ell equ}^{(1)\, e\mu uu} + G_S^{(d,p/n)}\, C_{\ell edq}^{e\mu dd} + G_S^{(s,p/n)} C_{\ell edq}^{e\mu ss}\,,\\
\widetilde C_{SR}^{(p/n)}&= \Big[\widetilde C_{SL}^{(p/n)} (e\leftrightarrow \mu)\Big]^*\,,
\end{align}
with the numerical coefficients~\cite{Kitano:2002mt}
\be
G_S^{(u,p)}=G_S^{(d,n)}=5.1\,,\quad 
G_S^{(u,n)}=G_S^{(d,p)}=4.3\,,\quad
G_S^{(s,p)}=G_S^{(s,n)}=2.5\,.
\ee
Notice that a more appropriate description of the the scalar form factors, using the effective Lagrangian at the nucleon level and including also the effective interaction with gluons, should be considered.
Nevertheless, the scalar operators do not enter our analysis and therefore we refer to Ref.~\cite{Cirigliano:2009bz} for further details.


\bibliographystyle{JHEP} 
\bibliography{refs}

\end{document}